\documentclass[12pt,a4paper]{article}

\usepackage[noblocks]{authblk}
\usepackage{a4wide,times}
\title{Bond Graph Modelling of Chemoelectrical Energy Transduction.}

\author[1,*]{Peter J. Gawthrop}
\author[2]{Ivo Siekmann}
\author[1]{Tatiana Kameneva}
\author[3]{Susmita Saha}
\author[3,4]{Michael R. Ibbotson}
\author[1,5,6]{Edmund J. Crampin}

\affil[1]{Department of Electrical and Electronic Engineering, The
  University of Melbourne, Australia.}
\affil[2]{Institute for Mathematical Stochastics, 
University of G\"{o}ttingen, Germany.}
\affil[3]{National Vision Research Institute, Australian College of
  Optometry, Australia.}
\affil[4]{Centre of Excellence for Integrative Brain Function,
  Dept. Optometry and Vision Sciences, The University of Melbourne,
  Australia.}
\affil[5]{School of Mathematics and Statistics, University of
      Melbourne University of Melbourne, Victoria 3010,
  Australia}
\affil[6]{School of Medicine, University of Melbourne, Victoria 3010,
  Australia}

\affil[*]{Corresponding author: \emph{peter.gawthrop@unimelb.edu.au}}

\usepackage{siunitx}

\usepackage{amsmath,amssymb,amscd,amstext,mathtools}
\newcommand{\mf}{\chi} 
\newcommand{\BG}[1]{\text{\sffamily\textbf{#1}}}

\newcommand{\C}{\BG{C }}

\newcommand{\R}{\BG{R }}

\newcommand{\one}{\BG{1 }}
\newcommand{\zero}{\BG{0 }}

\newcommand{\TF}{\BG{TF }}

\renewcommand{\Re}{\BG{Re }}

\newcommand{\BGL}[2]{$\BG{#1}$:$\mathbf{#2}$} 

\newcommand{\BC}[1]{\BGL{C}{#1}}

\newcommand{\BSS}[1]{\BGL{SS}{#1}}

\newcommand{\BTF}[1]{\BGL{TF}{#1}}

\newcommand{\BRe}[1]{\BGL{Re}{#1}}

\newcommand{\ddt}[1]{\frac{d#1}{dt}}
\newcommand{\VV}{\bar{V}}
\newcommand{\VT}{\tilde{V}}
\newcommand{\reacL}[2]{
  \xrightleftharpoons[#2]{#1} 
}
\newcommand{\reac}{
  \reacL{}{}
}
\newcommand{\reacu}[1]{
  \reacL{#1}{}
}
\newcommand{\lb}{\left (}
\newcommand{\rb}{\right )}

\newcommand{\K}{\text{K}^+}
\newcommand{\Na}{\text{Na}^+}
\newcommand{\Ca}{\text{Ca}^{2+}}

\usepackage[all]{xy}
\usepackage{array}

\newcommand{\bd}[1]{
\ar@_{->}[#1]
}
\newcommand{\bu}[1]{
\ar@^{->}[#1]
}
\newcommand{\cs}[1]{
\ar@{-}[#1]
}

\newcommand{\cc}[1]{
\ar@/^/[#1]
}

\usepackage[numbers,compress,sort]{natbib}
\usepackage{graphicx,subfigure,fancybox,color}

\newcommand{\Fig}[2]{
 \includegraphics[width=#2\linewidth]{#1.pdf}
  \label{subfig:#1}
}

\newcommand{\SubFig}[3]{
 \subfigure[#2]{
   \includegraphics[width=#3\linewidth]{#1.pdf}
   \label{subfig:#1}
 }
}

\usepackage[all]{xy}

\usepackage[pagebackref,hidelinks]{hyperref}
\usepackage{url}

\begin{document}
\maketitle

\abstract{
    Energy-based bond graph modelling of biomolecular systems is
    extended to include chemoelectrical transduction thus enabling
    integrated thermodynamically-compliant modelling of
    chemoelectrical systems in general and excitable membranes in
    particular.

    Our general approach is illustrated by recreating a well-known
    model of an excitable membrane. This model is used to investigate
    the energy consumed during a membrane action potential thus
    contributing to the current debate on the trade-off between the
    speed of an action potential event and energy consumption. The
    influx of $\Na$ is often taken as a proxy for energy consumption;
    in contrast, this paper presents an energy based model of action
    potentials. As the energy based approach avoids the assumptions
    underlying the proxy approach it can be directly used to compute
    energy consumption in both healthy and diseased neurons.

    These results are illustrated by comparing the energy consumption
    of healthy and degenerative retinal ganglion cells using both
    simulated and \emph{in vitro} data.
}

\newpage
\tableofcontents
\newpage
\section{INTRODUCTION}
\label{sec:introduction}
Energy is fundamental to science in general and the life sciences in
particular \citep{AtkPau11}. For this reason, the energy-based bond
graph modelling method, originally developed in the context of
engineering \citep{Pay61}, has been applied to modelling biomolecular
systems \citep{OstPerKat71,OstPerKat73,GawCra14,GawCurCra15,GawCra16,Gaw17}.
Bond graphs represent the energy consumption per unit time, or power
flow, for each component of a given system. Power flows are calculated
as the product of appropriately chosen `signal quantities' named
`efforts' and `flows'. Examples of effort include force, voltage,
pressure and chemical potential. Corresponding examples of flow are
velocity, current, fluid flow rate and molar flow rate. By describing
systems under the unifying principle of energy flows and due to the
abstract representation of these flows in terms of generalised effort
and flow quantities, the bond graph approach is particularly
appropriate for modelling systems with multiple energy domains. 
A comprehensive account of the use of bond graphs to model engineering
systems is given in the textbooks of \citet{GawSmi96},
\citet{MukKarSam06}, \citet{Bor11} and \citet{KarMarRos12} and a
tutorial introduction for control engineers is given by
\citet{GawBev07}. Chemical reactions are considered  by \citet{Cel91} and \citet{GreCel12}.
%
In particular, as an engineering example, \citet{Kar90} has looked at
chemoelectrical energy flows in hybrid vehicles.

Chemoelectrical energy transduction is fundamental to living systems
and occurs in a number of contexts including oxidative phosphorylation
and chemiosmosis, synaptic transmission and action potentials in
excitable membranes.
For this reason, this paper extends the energy-based bond graph
modelling of biomolecular systems to biological systems which involve
chemoelectrical energy transduction.  A simple model of the action
potential in excitable membranes is used as an illustrative example of
the general approach.
Although such simple models could easily be derived by other means,
our general approach could be used to build thermodynamically
compliant models of large hierarchical systems such as those describing
metabolism, signalling and neural transmission.

Understanding the biophysical processes which underlie the generation
of the action potential in excitable cells is of considerable
interest, and has been the subject of intensive mathematical and
computational modelling. Since the early work of \citet{HodHux52} on
modelling the ionic mechanisms which give rise to the action potential
in neurons,
mathematical models of the action potential 
have incorporated ever-increasing biophysical and ionic detail
\citep{Hill01}, and have been formulated to describe both normal and
pathophysiological mechanisms.
Generation of the action potential comes at a metabolic cost. Energy
is required to maintain the imbalance of ionic species across the
membrane, such that when ion channels open there is a flux of ions
(current) across the membrane -- initially carried by sodium ions --
generating rapid membrane depolarisation (the upstroke of the action
potential). 
Each action potential reduces the ionic imbalance and
each ionic species needs to be transported across the membrane against
an adverse electrochemical gradient to restore the imbalance -- this
requires energy.

The role of energy in neural systems has been widely discussed in the
literature
\citep{CloBolLow09,CarBea09,HasOttCal10,SenSteLau10,SenSteFri13,SenSte14,Niv16}
and it has been suggested that metabolic cost is a unifying principle
underlying the functional organisation and biophysical properties of neurons
\citep{NivLau08,HasOttCal10}.
Furthermore, \citet{Bea92} posed the question ``Does impairment of
energy metabolism result in excitotoxic neuronal death in
neurodegenerative illnesses?''  More recently, it has been suggested
\citep{Wel12,WelClo12,CloMidWel12,ReeSimDuc16} that an energy-based approach is
required to elucidate neurodegenerative diseases such as Parkinson's
disease.

In such studies, the flow of $\Na$ across the membrane is taken as a
proxy for energy consumption associated with action potential
generation, as $\Na$ has to be pumped back across the membrane by an
energy-consuming ATPase reaction.  This energetic cost is often quoted
as an equivalent number of ATP molecules required to restore the ionic
concentration gradient through activity of the sodium-potassium ATPase
(the Na pump), as calculated via stoichiometric arguments.
While this provides a useful indication of energetic cost, this is
however an imprecise approach, which cannot produce reliable estimates
of energy flows under all conditions (physiological and
pathophysiological). What is required instead is a way of simulating
and calculating the actual energy flows associated with these ionic
movements through a physically-based modelling approach.

Here we develop a general bond graph based modelling framework that
enables explicit calculation of the energy flows involved in moving
ions across the cell membrane. Our model can be applied to the
regulation of the membrane potential via ion channels in any cell
type---obvious examples include, for example, cardiac and skeletal
muscle cells---but in order to give a specific application we
investigate the energy cost for generating a neural action potential.
Because ions carry electrical charge, differences in the concentration
of a particular ionic species on either side of a membrane generate
both a chemical potential difference due to the concentration gradient
as well as an electrical potential difference due to charge
imbalance. 

The Hodgkin-Huxley model \citep{HodHux52} is used throughout this
paper as a well-known and well-documented exemplar to illustrate our
approach; however the approach is equally valid for more recent
models. In particular, we derive a bond graph formulation focusing on
the flow of energy associated with voltage-dependent ionic membrane
flow and this formulation encompasses both the classical
Hodgkin-Huxley model and more recent models.

One component of this formulation accounts for the behaviour of the
ion channel as an electrical resistor that can be investigated
experimentally via current-voltage relationships, also known as
$I$-$V$ curves. Popular models for~$I$-$V$ curves are, for example,
the linear~$I$-$V$ dependency that corresponds to Ohm's law and the
non-linear Goldman-Hodgkin-Katz (GHK) equations. Hodgkin-Huxley-like
models assume a linear~$I$-$V$ relationship.
  \citet{HodHux52} represented the voltage-dependent opening and
  closing of ion channels by an empirical model, their gating
  variables, which led to excellent agreement with experimental
  data. This paper replaces this empirical model of gating by an
  physically-based  voltage-dependent Markov model with one
  open and one closed state. Whereas this model accounts for the
  energy needed for moving the ion channel between conformations where
  the channel is in an open or a closed state, respectively, the
  empirical gating variables of the Hodgkin-Huxley model do not have a
  similar interpretation.

  As our models are formulated using bond graphs, important physical
  quantities such as mass and energy are conserved (in the sense that
  any dissipative processes are directly accounted for, and any inputs
  to or losses from the system are quantified).

  This theoretical approach is then applied to analyse energy
  consumption in retinal ganglion cells based on \emph{in vitro}
  experimental data collected and analysed from retinal ganglion cells
  (RGCs) of wild-type (WT) and degenerative (RD1) mice. Our use of the
  RD1 degenerate retina mouse model ensures that the outcomes of this
  project are directly relevant to human patients since RD1 mice have
  a degenerate retina that has distinct similarities to that observed
  in human patients with \emph{retinitis pigmentosa} -- a set of
  hereditary retinal diseases that results from the degenerative loss
  of the photoreceptors in the retina. 

A virtual reference environment \citep{HurBudCra14} is available for
this paper at \url{https://github.com/uomsystemsbiology/energetic_cost_reference_environment}.

\section{MODELLING APPROACH}
\begin{figure}[htb]
 \centering
 \Fig{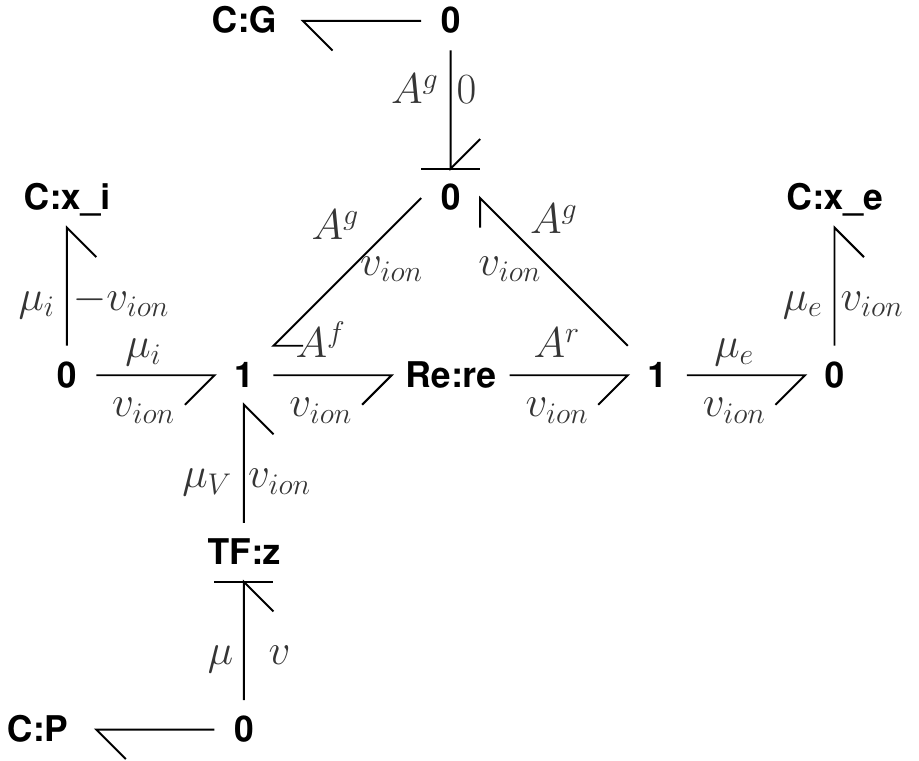}{0.5}
 \caption[Ion-channel Model]{
   Ion-channel Model: Reaction analogy. 
   With reference to reaction
   \eqref{eq:IC_reaction}, the ion $X_e$ external to
   the membrane is represented by \BC{x_e}, the ion $X_i$
   internal to the
   membrane is represented by \BC{x_i} and the internal to external
   molar flow rate is $v_{ion}$. \BTF{z} is the  electrostoichiometric
   transformer 
   where $z$ is the integer ionic charge. \BC{P} represents the
   membrane potential as an electrogenic capacitance with
   electrogenic potential $\mu$ and electrogenic flow $v$. The ionic
   flow is determined by the reaction component \BRe{re} and modulated
   by the gating affinity $A^g$ associated with the \BC{G}. 
   The two components \BC{G} and \BC{P} and associated junctions may
   be replaced by ports; this enables the model to be reused within a
   hierarchical framework.  }
 \label{fig:BG_channel}
\end{figure}

  We represent the flow of ions through the open pore of an ion
  channel in analogy to a chemical reaction:
\begin{equation}
  \label{eq:IC_reaction}
  X_i + zP + G \reac X_e + G
\end{equation}
The intracellular and extracellular concentrations of a particular ion
are represented as different chemical species, the intracellular
species $X_i$ and the extracellular species $X_e$. Conversion from
$X_i$ to $X_e$ and vice versa is modelled by an enzymatic reaction
with the \emph{gating species} $G$ that ``catalyses'' the flow of ions
across the cell membrane and accounts for the voltage-dependent
opening and closing of the channel as well as the behaviour of the
channel as an electrical resistor. The influence of the membrane
potential is represented in this reaction by the \emph{electrogenic
  species} $P$. Further below we will explain how an electrical
potential~$V$ can be converted to an equivalent chemical
potential~$\mu_V$.

A bond graph representation of~\eqref{eq:IC_reaction} according to the
framework developed by \citet{GawCra14} is given in
Figure~\ref{fig:BG_channel}. Using this specific example we briefly
review the bond graph approach developed in our previous work. The
components of a bond graph are distinguished based on how they
transform energy. Capacitors or springs store energy (\C), resistors
or dampers dissipate energy (\R), and transducers (or transformers)
(\TF) which transmit and convert, but do not dissipate, power. The
chemical species appearing in~\eqref{eq:IC_reaction} are represented
in the bond graph by capacitors \BC{x_i}, \BC{x_e}, \BC{P} and \BC{G}.
Because a given type of component usually occurs more than once in a
given system, the `colon' notation is adopted to distinguish between
different instances of each component type: the symbol preceding the
colon identifies the type of component, and the label following the
colon identifies the particular instance.

Let us first consider the species~$x_i$ and~$x_e$. These are simply
the concentrations of a particular ion within the cell~($x_i$) and
outside the cell~($x_e$) so that their rates of change must equal the
molar flow rate~$v_{ion}$ in \si{\mole\per\second} with opposite sign:
\begin{equation}
  \dot{x_i}=-v_{ion}, \quad \dot{x_e}=v_{ion}.
\end{equation}
It is assumed that in the body biochemical reactions occur under
conditions of constant pressure (isobaric) and constant temperature
(isothermal). Under these conditions, the chemical potential $\mu_A$
of substance $A$ measured in \si{\joule\per\mole} is given
\citep{AtkPau11} in terms of its mole fraction $\mf_A$ as:
\begin{equation} \label{eq:mu_A_0}
  \mu_A =  \mu_A^\star + RT \ln \mf_A 
\end{equation}
where $\mu_A^\star$ is the value of $\mu_A$ when $A$ is pure
($\mf_A=1$), $R = 8.314\si{~JK^{-1}mol^{-1}}$ is the universal gas
constant, $T~\si{K}$ is the absolute temperature and $\ln$ is the
natural (or Napierian) logarithm.
By introducing the \emph{thermodynamic constant} $K_A$
\begin{equation}
  \label{eq:K_A}
  K_A = \frac{1}{n_{total}}\exp{\frac{\mu_A^\star}{RT}} 
\end{equation}
where $n_{total}$ is the total number of moles in the mixture we can express the chemical potential as a function of molar amount $x_A$ of the species~$A$:
%
\begin{equation}\label{eq:C_chem}
 \mu_A = RT \ln K_A x_A. 
\end{equation}
As discussed by \citet{OstPerKat71,OstPerKat73} the 
chemical potential 
$\mu$ and the molar flow~$v$ are appropriate \emph{effort} and
\emph{flow} variables for modelling chemical reactions. The product of
chemical potential~$\mu$ and molar flow~$v$ is the energy flow into
the bond graph \C components and has the unit \si{\joule\per\second}
of \emph{power}. This is shown in Figure~\ref{fig:BG_channel} by power
bonds (or more simply `bonds'), drawn as harpoons:
$\rightharpoondown$.
These bonds can optionally be annotated with specific effort and flow
variables, for example
$\xrightharpoondown[f]{e}$.
The half-arrow on the bond indicates the direction in which power
power flow will be regarded as positive and thus defines a sign
convention.  As explained above, for the components \BC{x_i} and
\BC{x_e} we have the flows $-v_{ion}$ and $v_{ion}$, respectively, and
the chemical potentials are \citep{GawCra14}:
\begin{align}
 \mu_i &= RT \ln K_i x_i \label{eq:mu_i}\\
 \mu_e &= RT \ln K_e x_e \label{eq:mu_e}\\
\text{where } K_i &= \frac{K_{ion}}{C_i}\\
\text{and } K_e &= \frac{K_{ion}}{C_e}\\
\text{where } K_{ion} &= \exp \frac{\mu^0}{RT}
\end{align}
$\mu^0$ is the standard chemical potential for the ion and the
volumetric capacities of the interior and exterior are $C_i$ and
$C_e$, respectively. 
In general, bond graph components which are connected by a bond share
the same effort and flow. Thus, because all energy flows between bond
graph components are quantified, the bond graph ensures that
properties such as energy and mass are conserved.

We follow \citet{OstPerKat73} in describing chemical reactions in
terms of the \emph{Marcelin -- de Donder} formulae as discussed by
\citet{Rys58} and \citet{GawCra14}. In particular, given the $i$th
reaction \citep[(5.9)]{OstPerKat73}:
\begin{equation}
  \label{eq:react}
  \nu^f_A A + \nu^f_B B + \nu^f_C C  \dots 
  \reacu{i} 
  \nu^r_A A + \nu^r_B B + \nu^r_C C \dots 
\end{equation}
where the stoichiometric coefficients $\nu$ are either zero or
positive integers and the \emph{forward affinity} $A^f_i$ and the
\emph{reverse affinity} $A^r_i$ are defined as:
\begin{align}
  A^f_i &= \nu^f_A \mu_A + \nu^f_B \mu_B + \nu^f_C \mu_C \dots  \label{eq:A^f}\\
  A^r_i &= \nu^r_A \mu_A + \nu^r_B \mu_B + \nu^r_C \mu_C \dots  \label{eq:A^r}
\end{align}
  Thus, the affinities~$A^f_i$ and~$A^r_i$ are sums of the chemical
  potentials appearing on either side of~\eqref{eq:react} weighted
  with the stoichiometric coefficients~$\nu$. In a bond graph model,
  the affinities from~\eqref{eq:A^f}, \eqref{eq:A^r} can be obtained
  via a `\one junction'. Junctions allow parallel (`\zero' junction)
  and series (`\one' junction) connections to be made. The efforts on
  bonds impinging on a \one junction sum to zero whereas the flows are
  all equal (this is equivalent to Kirchhoff's second law for
  electrical circuits, where effort is equivalent to voltage and flow
  is equivalent to current). Analogously, for a \zero junction the
  flows sum to zero and the efforts are all equal (equivalent to
  Kirchhoff's first law). Taking into account the sign convention
  explained above it is easy to see that~$A^f$ and~$A^r$ in
  Figure~\ref{fig:BG_channel} are indeed weighted sums of chemical
  potentials as in~\eqref{eq:A^f}, \eqref{eq:A^r}:
%
\begin{align}
 A^f &=  A^g(V) + \mu_i + \mu_V 
       \label{eq:Af}\\
 A^r &=  A^g(V) + \mu_e 
\end{align}
The affinity of the gating species~$A^g(V)$ as well as the
potential~$\mu_V$ of the electrogenic species will be derived further
below.

The Marcelin -- de Donder formulae enable us to calculate the flow of
a reaction for which we assume the law of mass action from the
affinities:
\begin{align}
 v_i &=  \kappa_i \lb v^+_0 - v^-_0 \rb \label{eq:v_exp}\\
   \text{where }
   v^+_0 &=   e^\frac{A^f_i}{RT}
   \text{ and }
   v^-_0 =  e^\frac{A^r_i}{RT} 
\end{align}
Note that the
arguments of the exponential terms are dimensionless as are $v^+_0$
and $v^-_0$. The units of the reaction rate constant $\kappa_i$ are
those of molar flow rate: $\si{mol.sec^{-1}}$.

  The flow~$v_i$ is obtained from the efforts~$A^f_i$ and~$A^r_i$,
  similar to an electrical resistor where Ohm's law requires that the
  current is proportional to the potential difference across the
  resistor. In contrast to the linear electrical \R component, the
  non-linear effort-flow relationship of~\eqref{eq:v_exp} can be
  represented by a two port resistive component, the \Re component
  \citep{GawCra14}. Resistive components such as \R and \Re components
  dissipate energy.

  So far we have not accounted for the interdependency of the molar
  flow~$v_{ion}$ and the electrical potential~$V$ across the cell
  membrane. As discussed in bond graphs terms by \cite{Kar90} the
  well-known physical relationship between the molar flow of
  ions~$v_{ion}$ across the membrane and the current~$i$
\begin{equation}
 \label{eq:faraday}
   i = zF v_{ion}
\end{equation}
in fact describes the coupling of two energy domains, namely chemical
energy and electrical energy.
Here $z$ is the (integer) ionic charge and $F$ Faraday's constant
which has the approximate value
$F \approx 96.5\times 10^{3}\text{C mol}^{-1}$ Thus, for example, an
ion with unit charge and flow rate of $1~\si{nmol.sec^{-1}}$ is
equivalent to a current of about $96.5 \mu~\si{A}$.
\begin{figure}[htbp]
 \centering
 \Fig{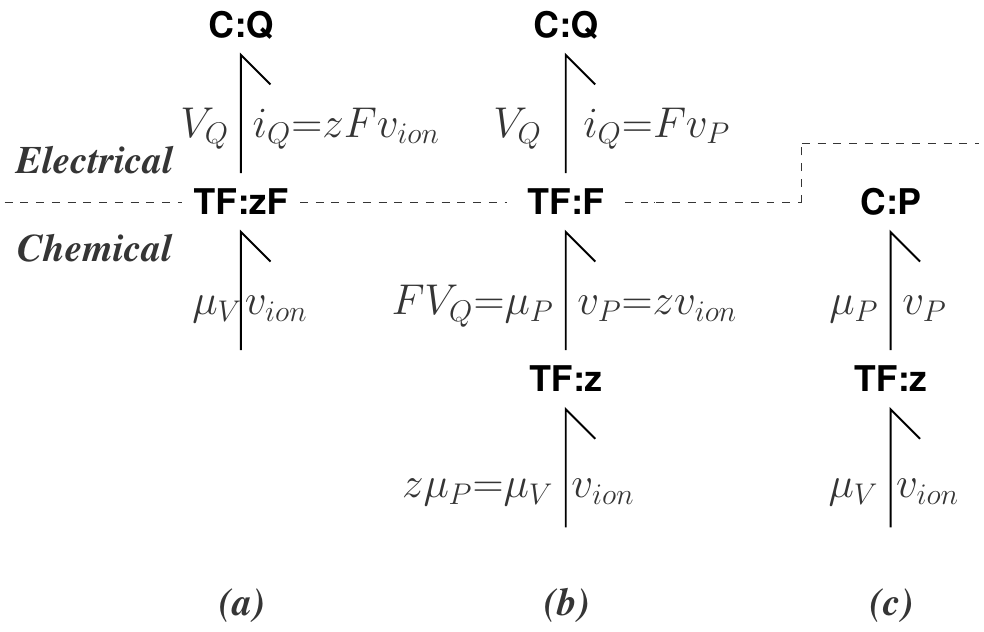}{0.6}
 \caption[Chemoelectrical Transduction.]{Chemoelectrical
   Transduction. 
   (a) The transduction of ion membrane flow $v_{ion}$ into electrical
   current $i$ is represented by \BTF{zF} where where $z$ is the
   ionic charge and $F$ Faraday's constant; \BC{Q} represents the
   membrane capacitance containing electrical charge $Q$ and $\mu_V$ is the \emph{chemoelectrical potential}: the
chemical potential corresponding to the membrane voltage $V$ 
   (b) \BTF{zF} is split into two series components \BTF{z} \&
   \BTF{F}. 
   (c) The electrical capacitor \BC{Q} and the chemoelectrical
   transducer \BTF{F} are combined into chemo-equivalent capacitor
   \BC{P} containing $x_m$ moles of charge. \BC{P} is analogous to
   a chemical species, and \BTF{z} to a stoichiometric transformer.  }
 \label{fig:Cequiv}
\end{figure}

  Using the bond graph \emph{transformer} element denoted by ~\TF,
  equation~\eqref{eq:faraday} enables us to associate a
  \emph{chemoelectrical potential} $\mu_V$ with an ionic current~$i$
  at a membrane potential~$V$. As shown in Figure~\ref{fig:Cequiv}(a), the transformer~\BTF{zF} converts electrical power~$V_Q i$ to chemical power~$\mu_V v_{ion}$ so that from $V_Q i= \mu_V v_{ion}$ we can derive the defining equation of the chemoelectrical potential
\begin{equation}
 \label{eq:faraday_e}
 \mu_V = zF V_Q
\end{equation}
  Thus, for example, a voltage of $1mV$ is equivalent to a chemical
  potential of an ion with unit charge of about $96.5 \si{~J.mol^{-1}}$
%

Stoichiometric analysis of biochemical reactions has a bond
graph interpretation in terms of the bond graph structure. 
As a structural matrix, the stoichiometric matrix for biochemical systems contains integer entries
corresponding to reaction stoichiometry.
%
The bond graph describing chemoelectrical transduction in
Figure \ref{fig:Cequiv}(a) contains transformers connecting the
electrical and chemical domains and the chemoelectrical transformer 
contains the Faraday constant $F$ which is not an integer. 
For this reason, it is useful as shown in Figure \ref{fig:Cequiv}(b) to split the component
\BTF{zF} into two \TF components in series: \BTF{z} and
\BTF{F}. Thus \BTF{z} corresponds to the integer, but ion-dependent
charge, $z$, and \BTF{F} to the universal constant, but non-integer,
$F$.
\BTF{z} will be referred to as the \emph{electrostoichiometric
 transformer} with ratio $z$. With reference to Figures \ref{fig:Cequiv}(b)\&(c)
\begin{xalignat}{2}
  v &= zv_{ion}& \mu_V &= z\mu_Q
\end{xalignat}

Furthermore, the electrical component \BC{Q} and the chemoelectrical
transformer \BTF{F} component may be replaced by a single \emph{electrogenic
capacitor} \BC{P} as in Figure \ref{fig:Cequiv}(c). 
Assuming that the electrical capacitor
has the constant capacitance $C~\text{(CV}^{-1})$,
and with reference to Figure
\ref{fig:Cequiv}(b)
\begin{xalignat}{4}\label{eq:C_actual}
 \mu_Q &= F V_Q &
 V_Q &= \frac{Q}{C}&
 \dot{Q} & = i_Q&
 i_Q &= Fv_P
\end{xalignat}
%
%
While the electrogenic capacitor \BC{P} is linear and therefore cannot
be written in the logarithmic form of Equation \eqref{eq:C_chem}, it
is convenient to write the defining equations in a similar form by
defining the number of moles of charge $x_m$ as
\begin{equation}
  x_m = \frac{Q}{F}
\end{equation}
With reference to Figure \ref{fig:Cequiv}(c)
\begin{xalignat}{2}\label{eq:C_equiv}
 \mu_P &= RT K_e x_m &
 \dot{x}_m & = v_P
\end{xalignat}
Comparing Equations \eqref{eq:C_actual} and \eqref{eq:C_equiv}, it
follows that:
\begin{align}\label{eq:V_N}
 K_e &= \frac{F^2}{CRT} = \frac{V_e}{V_N}\text{mol}^{-1} \\
 \text{where } V_e &= \frac{F}{C} \text{Vmol}^{-1} 
 \text{ and } V_N = \frac{RT}{F} \text{V} \notag
\end{align}
$V_e$ is the \emph{equivalent voltage} associated with each mole of
charge and is dependent on the electrical capacitance $C$. $V_N$ is
related to the \emph{Nernst potential} and is temperature
dependent.

In summary, \BC{P} is analogous to a chemical species (but with a non-logarithmic
characteristic), and \BTF{z} is analogous to the stoichiometric
transformers discussed by \citet{OstPerKat71,OstPerKat73},
\citet{GreCel12} and \citet{GawCra14}.
\begin{figure}
  \centering
  \Fig{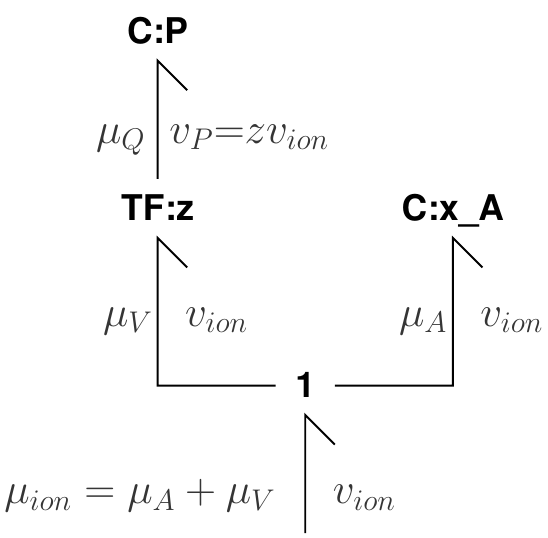}{0.3}
  \caption{Combined chemoelectrical and chemical potentials}
  \label{fig:CC}
\end{figure}
  Because the influence of the membrane potential~$V$ can be
  translated to the contribution of the electrogenic species, its
  effect on the flow of a species~$A$ can be accounted for in the bond
  graph just like a reaction of two normal species by coupling the
  efforts $\mu_V$ and $\mu_A$ via a \one junction \citep{GawCra14} as
  shown in Figure~\ref{fig:CC}. From the bond graph representation we
  rederive the standard electrostatic contribution to the
  chemoelectrical potential:
  \begin{equation}
 \label{eq:faraday_e_1}
 \mu_{ion} = \mu_A + \mu_V = \mu_A +  z\mu_Q = \mu_A +  zF V_Q = \mu_A +  F V
\end{equation}
In general, as discussed below, the electrogenic capacitor will
correspond to the net flow of more than one charged species.

  Similar to our approach to modelling the membrane potential we will
  also represent the voltage-dependent gating of ion channels as a
  chemical species. In this way we obtain a general model for ionic
  flow through ion channels across cell membranes. We will demonstrate
  that our general model of voltage-dependent gating includes the
  original Hodgkin-Huxley model and its extensions beyond
  sodium~$\text{Na}^+$, potassium~$\text{K}^+$ and leak currents.  As
  a specific application of our framework we will investigate the
  energy flow related to the action potential predicted by
  Hodgkin-Huxley like models.  Our analysis clarifies that the
  phenomenological gating variables introduced by \cite{HodHux52} fail
  to account appropriately for energy flows related to ion channel
  gating. 

  As shown in the chemical reaction~\eqref{eq:IC_reaction} that our
  model is based upon, the \emph{gating species} $G$ behaves like an
  enzyme that catalyses the conversion of the intracellular
  species~$X_i$ to the extracellular species~$X_e$. 
  In the same way as in \citet[Figure 2e]{GawCra14} 
  the \emph{gating
    affinity} $A^g$ is added to both sides of the reaction (Figure
  \ref{fig:BG_channel}). The second port \BSS{[g]} imposes the gating
  affinity $A^g$ and the corresponding flow
  $v_g=v_{ion}-v_{ion}=0$
  . With the gating affinity
\begin{align}
 A^g &= \mathbf{R T} \ln G_{pore} (V) + \mathbf{R T} \ln G_{ion}(V)   \label{eq:Ag}\\
\text{or }  \exp (A^g/\mathbf{R T}) &=  G_{pore} (V) G_{ion}(V) \label{eq:expAg}
\end{align}
we represent two characteristics of the ionic flow through a
channel. The term $G_{pore}(V)$ accounts for the fact that an ion
channel has an electrical resistance that opposes the ionic
current. The electrical resistance is commonly investigated
experimentally by determining current-voltage relationships. It will
be demonstrated that the two most commonly used models for
current-voltage relationships of ion channels, Ohm's law and the
Goldman-Hodgkin-Katz (GHK) equations, can be obtained by suitable
choices of $G_{pore}(V)$. Whereas $G_{pore}(V)$ represents the
conductance through an open channel, $G_{ion}(V)$ provides a model for
the voltage-dependent opening and closing of the channel.
From Equations (\ref{eq:A^f}-\ref{eq:v_exp}) we obtain our model for
the ionic flow through an ion channel:
\begin{align}
  v_{ion} &= \kappa \left[ \exp{ \left( \frac{A^g(V)+\mu_i+\mu_V}{RT} \right) }  - \exp{ \left( \frac{A^g(V)+\mu_e}{RT} \right) } \right] \\
&  = \kappa K_{ion} \exp{A^g(V
)} \left (  c_i \exp \VV - c_e \right )  \label{eq:v_1}\\
\text{where }  
c_i &= \frac{x_i}{C_i},\; c_e = \frac{x_e}{C_e}, \; V_N=\frac{R T}{F} \text{ and } \VV:=\frac{V}{V_N} 
\end{align}
%
Define $\VV_{ion}$ as the voltage for which $v_{ion}$ of Equation (\ref{eq:v_1})
is zero:
\begin{align}
 c_i \exp \VV_{ion} - c_e  &= 0 \label{eq:v_1_0}\\
 \text{hence } 
 \VV_{ion} &= \ln \frac{c_e}{c_i} = -\ln \frac{c_i}{c_e}\label{eq:VV_0}
\end{align}
Using Equations (\ref{eq:VV_0}) and \eqref{eq:expAg}, Equation \eqref{eq:v_1}  becomes:
\begin{align}
 v_{ion} &= \kappa K_{ion} c_e G_{ion}(V)G_{pore} (V) \left ( \exp \VT_{ion} - 1  \right
 )  \label{eq:v_mass-action}
\end{align}
where $\VT_{ion}$ is given by
\begin{equation}
  \label{eq:VT}
  \VT_{ion}=\frac{V-V_{ion}}{V_N} = \VV - \VV_{ion}.
\end{equation}
The quantity $\exp \left ( -\VT_{ion} \right )$ is known as the
  \emph{Ussing Flux Ratio} \cite[\S 3.2]{KeeSne09}. 
  The ionic flow (ion current) model of Equation
  \eqref{eq:v_mass-action} will be used in the sequel to construct a
  Bond Graph model from the Hodgkin Huxley model.

\begin{figure}[htbp]
 \centering
 \Fig{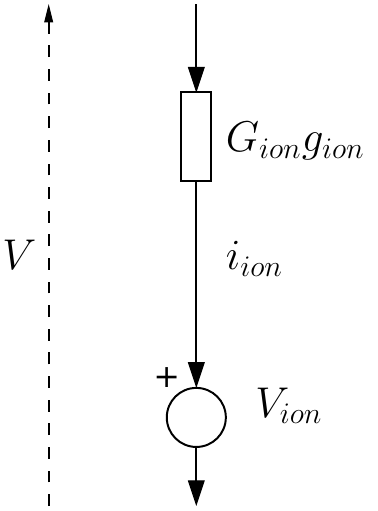}{0.2}
 \caption{The Hodgkin-Huxley Axon Model}
 \label{fig:HH}
\end{figure}

%
\citet{HodHux52} model an ion channel according to Ohm's law i.e. as a
linear conductance $g_{ion}$, modulated by a function $G_{ion}$ in
series with the Nernst potential represented by a voltage source
$V_{ion}$ (see Figure \ref{fig:HH}
). 

\begin{equation}\label{eq:HHi}
 i_{ion} = g_{ion} G_{ion}(V,t) (V - V_{ion})
\end{equation}
where $0 \le G_{ion}(V,t) \le 1$ is the gating function, which is a
dynamic function of membrane potential $V$. 
In terms of ionic flow $v_{ion}$, Equation \eqref{eq:HHi} becomes, using \eqref{eq:VT}:
\begin{align}
 v_{ion} = \frac{1}{F} i_{ion} &= \frac{g_{ion} G_{ion}(V,t) V_N}{F}
 \VT_{ion} \notag\\
 &= \kappa_{HH} c_e G_{ion} (V,t) \VT_{ion}\label{eq:HHv}\\
 \text{where } \kappa_{HH} &= \frac{g_{ion}V_N}{Fc_e}
\end{align}
It is easy to see that our model~\eqref{eq:v_mass-action} contains the
Hodgkin-Huxley model~\eqref{eq:HHv} by a suitable choice of
$G_{pore}$:
\begin{equation}
G_{pore} = G_{HH} (V) =
\begin{cases}
 1 & \VT = 0\\
 \frac{\VT_{ion}}{\exp \VT_{ion} - 1} & \VT_{ion} \ne 0
\end{cases}
\label{eq:G_HH}
\end{equation}
Clearly, equation (\ref{eq:G_HH}) only makes sense if $G_{HH}$ is
positive for all $\VT$. If $\VT<0$, both $\VT$ and $\exp \VT - 1$ are
negative; if $\VT>0$, both $\VT$ and $\exp \VT - 1$ are positive; and,
as $G_{HH}(0)=1$, $G_{HH}(\VT)$ is positive for for all $\VT$. 

A number of alternative physically-based models for the ion channel are available. In
particular, the Goldman-Hodgkin-Katz (GHK) model (see \citet[\S~2.6.3
(2.123)]{KeeSne09}, \citet[\S~9.1.1]{Koc04}) \& \citet{SteGraGil11}
can be rewritten in a similar form to (\ref{eq:v_mass-action}) as:
\begin{equation}
 \label{eq:GHK}
 v = \kappa K_0 c_e \VV \frac{\exp \VT - 1}{\exp \VV - 1}
\end{equation}
Comparing Equations (\ref{eq:v_mass-action}) and (\ref{eq:GHK}), it follows that
the mass action model \eqref{eq:v_mass-action} and GHK model are the
same if the model-dependent function $G_{pore} (V)$ is:
\begin{equation}
G_{pore} = G_{GHK} (V) =
\begin{cases}
 1 & \VV = 1\\
 \frac{\VV}{\exp \VV - 1} & \VV \ne 1
\end{cases}
\label{eq:G_GHK}
\end{equation}
Note that $G_{GHK} (V)$ \eqref{eq:G_GHK} is of the same form as
$G_{HH} (V)$ \eqref{eq:G_HH} except that $\VT$ is replaced by $\VV$.

\begin{figure}[htbp]
 \centering
 \SubFig{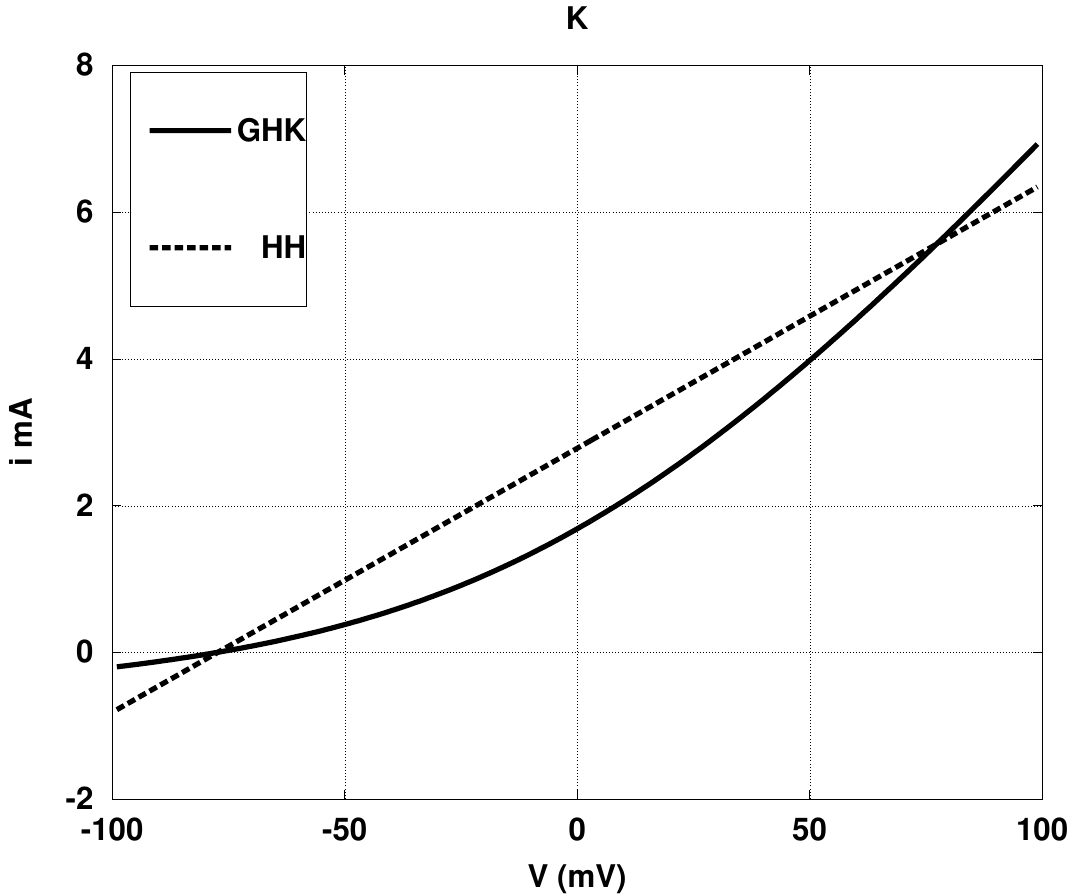}{$\text{K}^+$: current (\si{mA})}{0.45}
 \SubFig{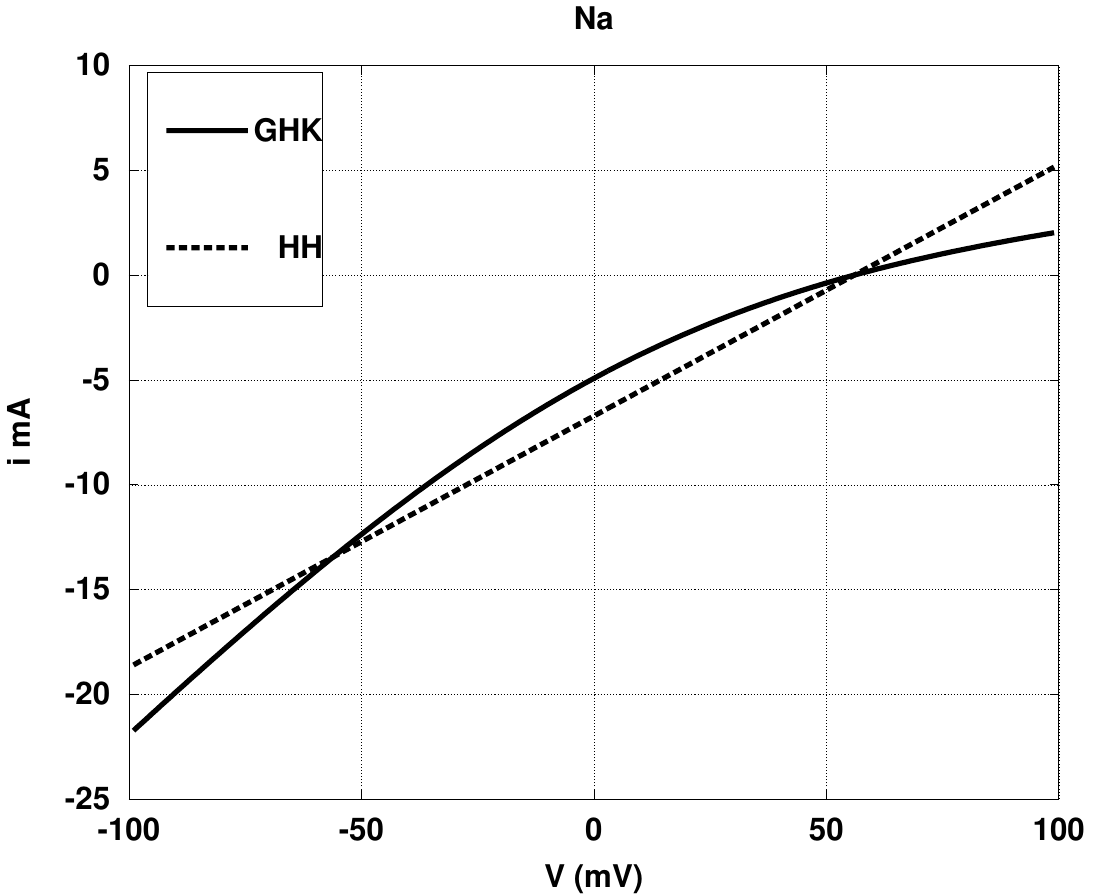}{$\text{Na}^+$: current (\si{mA})}{0.45}
 \caption{Comparison of current(\si{mA})--voltage(\si{\milli V}) relationships for
   $\text{Na}^+$ and $\text{K}^+$ currents  for $G_{pore}=G_{HH}$ and $G_{pore}=G_{GHK}$
   models.}
 \label{fig:comparison}
\end{figure}

From equations \eqref{eq:HHv} and \eqref{eq:v_mass-action}, both the
HH and GHK ion channel models give zero ionic flow when the membrane
voltage equals the Nernst voltage: that is the models match at
$\VT_{ion}=0$. Moreover, the GHK model of Equation \eqref{eq:GHK} has
a parameter $\kappa$ that can be chosen to fit the data. In this case,
$\kappa$ is chosen so that the GHK and HH models also match at another
voltage; in this case chosen as minus the Nernst voltage.
Figure \ref{fig:comparison} shows the ionic currents plotted against membrane
voltage
%
%
for the  $\text{K}^+$ and $\text{Na}^+$ channels and they match at the two
voltages. The GHK model is used in the sequel. 

\begin{table}[htbp]
 \centering
 \begin{tabular}{|l||l|l|l|}
   \hline
   Ion & $\K$ & $\Na$ & Leakage\\
   \hline
   $\kappa\si{~nmol.sec^{-1}}$ & $0.046262$ & $0.13204$ & $0.0014329$\\
   \hline
 \end{tabular}
 \caption{GHK parameter $\kappa \si{~nmol.sec^{-1}}$}
 \label{tab:GHK}
\end{table}

\label{sec:mass-action-model}

\section{Bond Graph Modelling of Voltage Gating}\label{sec:VoltageGating}
\begin{figure}[htbp]
 \centering
 \SubFig{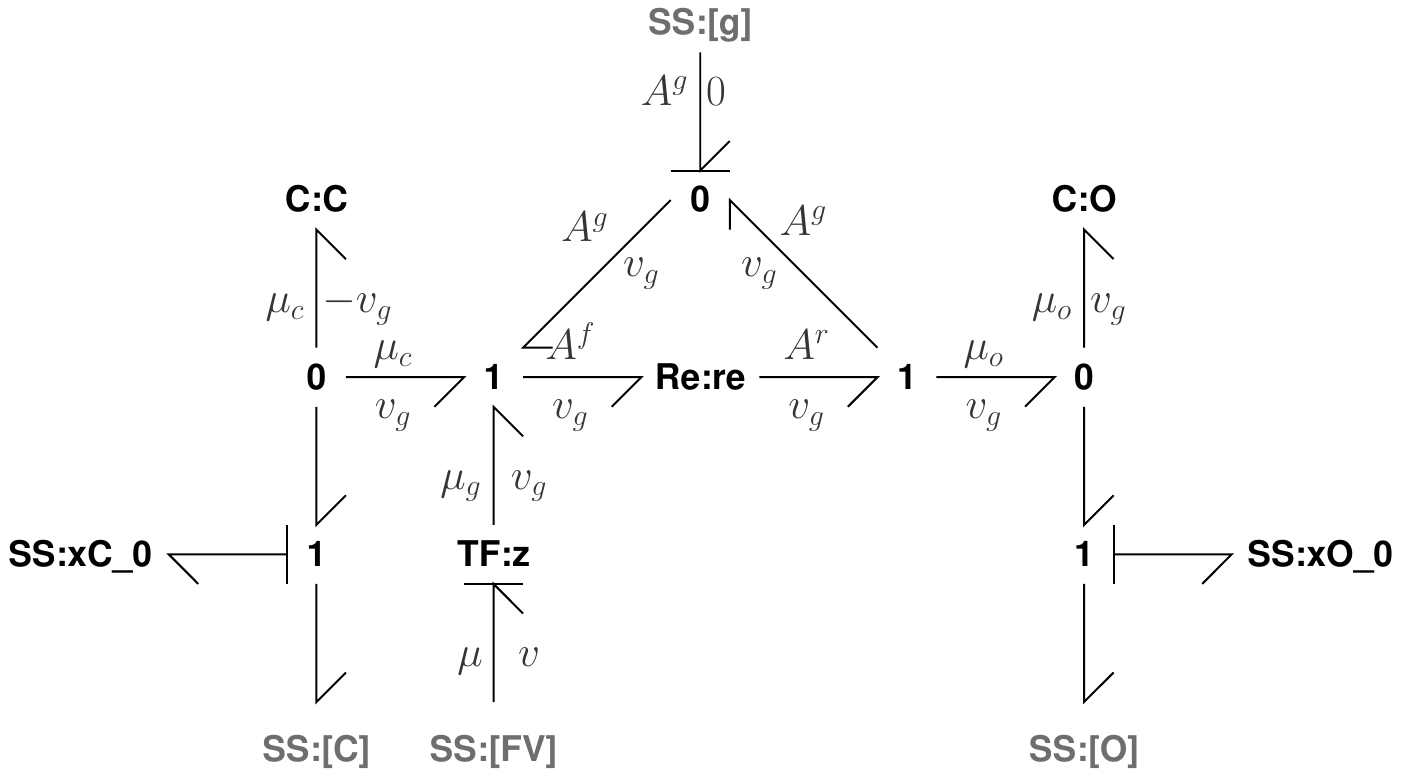}{Gate}{0.6}
 \SubFig{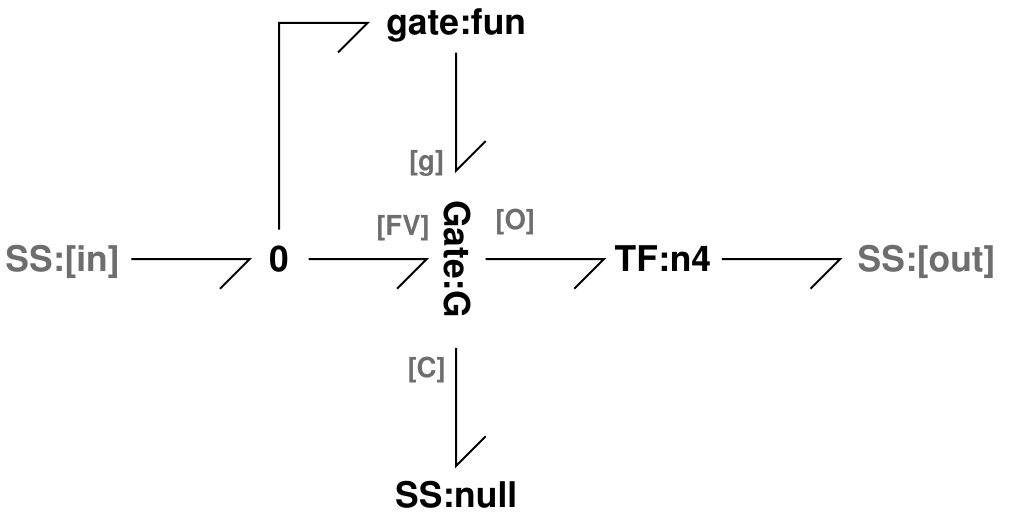}{GateK}{0.45}
 \SubFig{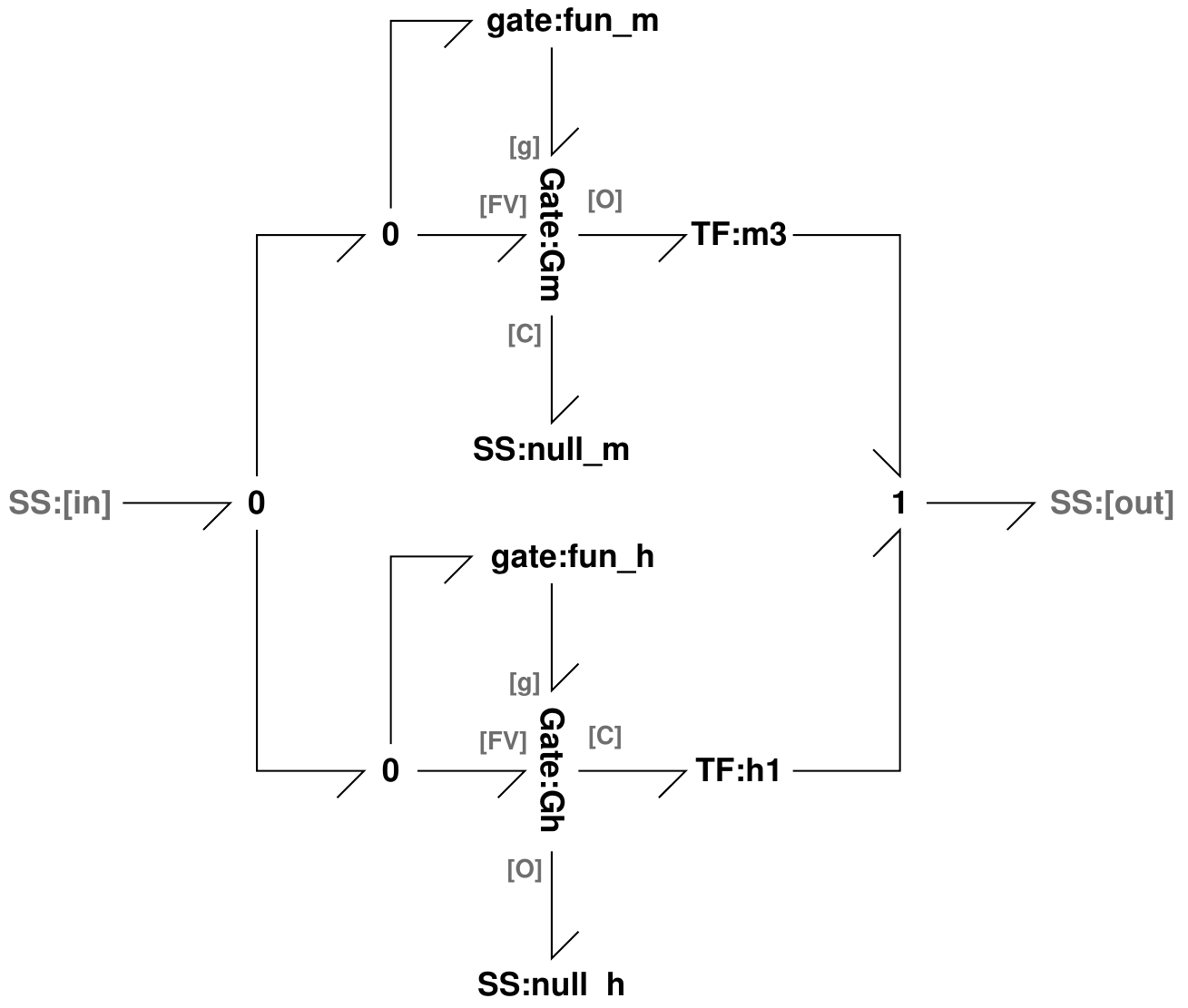}{GateNa}{0.45}
 \caption{Bond Graph model for physically-plausible gating. 
   (a) The voltage-modulated gate is modelled in a similar way the
   the ion channel of Figure \ref{fig:BG_channel}.
   (b) The basic gate of (a) is used in the $\K$ gate. The open
   state is used and the closed state discarded. The gate equations
   \eqref{eq:gate_ode_0} are implemented in \textbf{gate} and the
   $n^4$ factor 
   by \BTF{n4}.
   (c) The basic gate of (a) is used in the $\Na$ gate. There are
   separate models for the $m$ and $h$ gates. The
   $m^3$ factor 
   by \BTF{m3}.
}
 \label{fig:Pgate_abg}
\end{figure}

  Up to this point our general energy-based framework for ionic
  transport is consistent with the Hodgkin-Huxley (HH) model. However,
  our energy-based approach and the HH model differ in the description
  of voltage-dependent gating. With the gating function~$G_{ion}$ we
  represent the proportion of open channels. As explained by
  \citet{KeeSne09}, $G_{ion}$ may depend on one or more so-called
  gating variables which can be interpreted mechanistically as a model
  for an ion channel consisting of multiple subunits or gates. The
  subunits are assumed to open and close independently and for the
  channel to be open all subunits must be in an open state. According
  to this interpretation, the~$\text{K}^+$ channel of the HH model
  with
  \begin{equation}
    \label{eq:Kplus}
    G_{\text{K}^+} = n^4
  \end{equation}
  consists of four identical subunits whereas the~$\text{Na}^+$
  channel with
\begin{equation}
  \label{eq:Naplus}
  G_{\text{Na}^+} = m^3 h
\end{equation}
consists of three subunits of one and a single subunit of another
type. 
The gating variables are all represented by linear first-order differential equations with voltage-dependent coefficients of the form:
\begin{equation}
 \label{eq:gate_ode_0}
 \ddt{g} = \alpha(V) (1-g) -\beta(V) g
\end{equation}
This can be rewritten in a more convenient form as:
\begin{align}
 \ddt{g} &= -\tau(V) \lb g_{ss}(V) - g\rb \label{eq:gate_ode}\\
\text{where } \tau(V) &= \frac{1}{\alpha(V) + \beta(V)}\label{eq:tau}\\
\text{and } g_{ss}(V) &= \frac{\alpha(V)}{\alpha(V) + \beta(V)}\label{eq:g_inf}
\end{align}
  By interpreting~\eqref{eq:gate_ode_0} as the equation for the open
  probability of a two-state Markov model with one open and one
  closed state, the gating variable~$g$ can be regarded---by
  averaging over many gates---as the proportion of open gates in a
  population of gates.

  In the context of our energy-based framework we now have to
  determine under which circumstances voltage-dependent two-state
  Markov models of the form~\eqref{eq:gate_ode_0} that are used for
  each gate appearing in the HH model can be given a physical
  interpretation that enables us to keep track of the energy
  dissipated by the movement of ions via opening and closing of the
  gate.
As discussed by \citet[\S 3.5]{KeeSne09} and \citet[Chapter
2]{Hill01}, it is possible to represent ion channel gates as a set of
chemical equations incorporating \emph{gating charge} and \emph{gating
  current}.
  This means that the voltage-dependent transitions of a number of
  gates
\begin{equation}
 \label{eq:x_g}
 x_g = x_c + x_o
\end{equation}
between open and closed states are described as the first-order
reaction
\begin{equation}
 \label{eq:C2O}
 C \reacL{V}{} O.
\end{equation}
Because~\eqref{eq:C2O} has the same structure as the chemical
reaction~\eqref{eq:IC_reaction} we can derive our model for
voltage-dependent gating from the bond graph shown in
Figure~\ref{fig:Pgate_abg}(a) which is completely analogous to
Figure~\ref{fig:BG_channel}. From the bond graph of
Figure~\ref{fig:Pgate_abg}(a) we obtain the molar flow
\begin{align}
 v_g &= \frac{1}{x_g}\kappa(V) \lb e^{\frac{V}{V_g}} k_c x_c - k_o x_o \rb\notag\\
 &= \kappa(V) \lb e^{\frac{V}{V_g}} k_c \lb 1-\frac{x_o}{x_g} \rb - k_o \frac{x_o}{x_g} \rb \label{eq:x_o}
\end{align}
By comparison with the model for the gating variables this shows that
the reason why the HH model is not thermodynamically compliant is
that we cannot choose the voltage-dependent rates~$\alpha(V)$
and~$\beta(V)$ arbitrarily. In order to compare our approach with the HH model we choose the parameters~$V_g$, $k_c$, $k_o$ and the voltage-dependent rate constants~$\kappa(V)$ for each gate so that it mimics the HH model as closely as possible. 
This is illustrated for the three gates (n,m,h) used by \citet{HodHux52} as
listed by \citet{KeeSne09}, Equations 5.24--5.29, where the Bond Graph parameters have been chosen to fit the empirical HH  model of the form of Equation \eqref{eq:gate_ode_0} or \eqref{eq:gate_ode}. 
Figures \ref{subfig:Pgate_n_g}, \ref{subfig:Pgate_m_g} and
\ref{subfig:Pgate_h_g} show $g_{ss}$ for each of the three gates and
for both the physical and empirical values.
\begin{figure}[htbp]
 \centering
 \SubFig{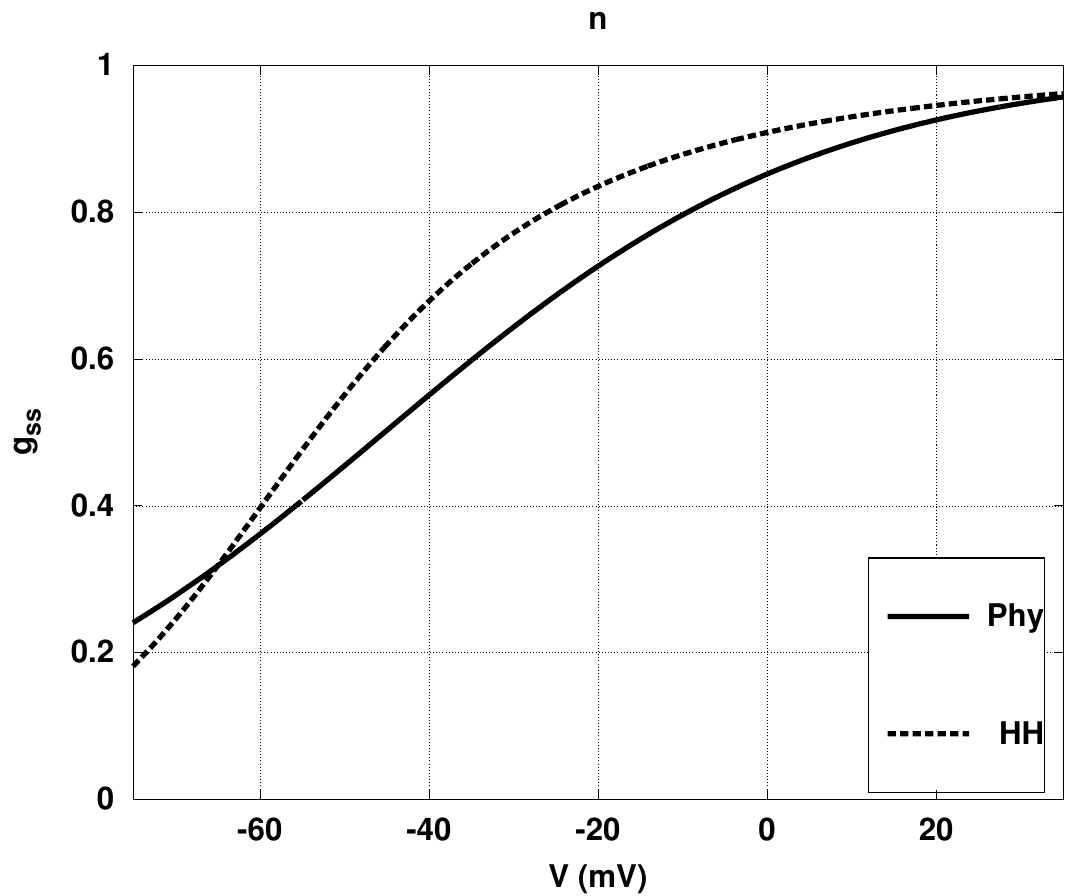}{n-gate $g_{ss}$}{0.45}
 \SubFig{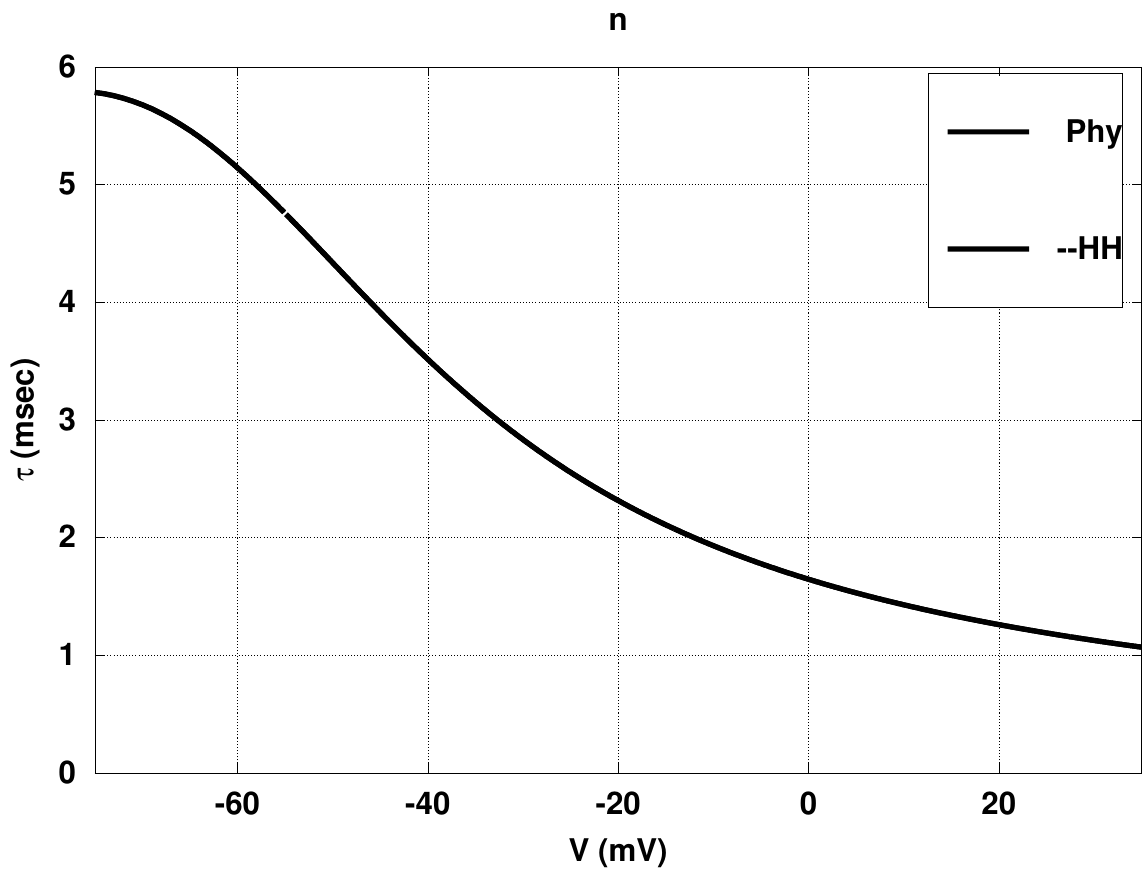}{n-gate $\tau$}{0.45}
%
 \SubFig{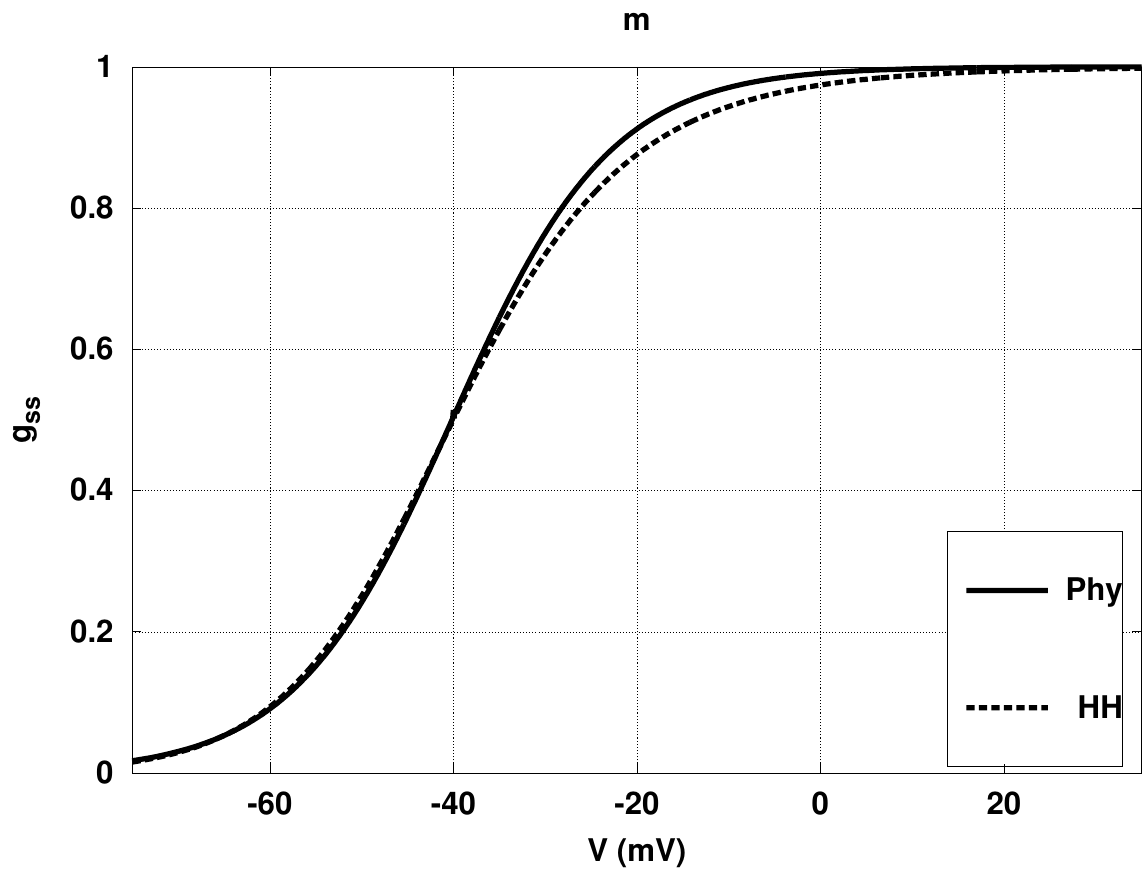}{m-gate $g_{ss}$}{0.45}
 \SubFig{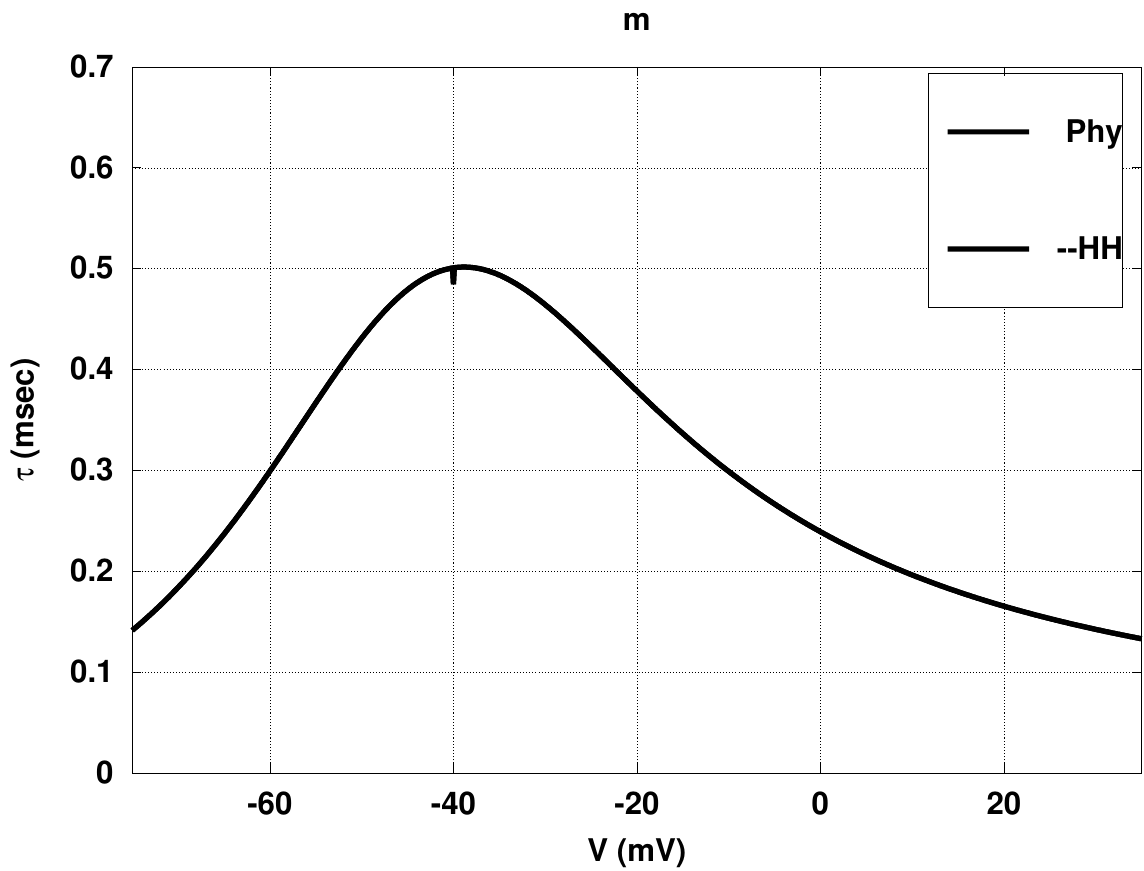}{m-gate $\tau$}{0.45}
%
 \SubFig{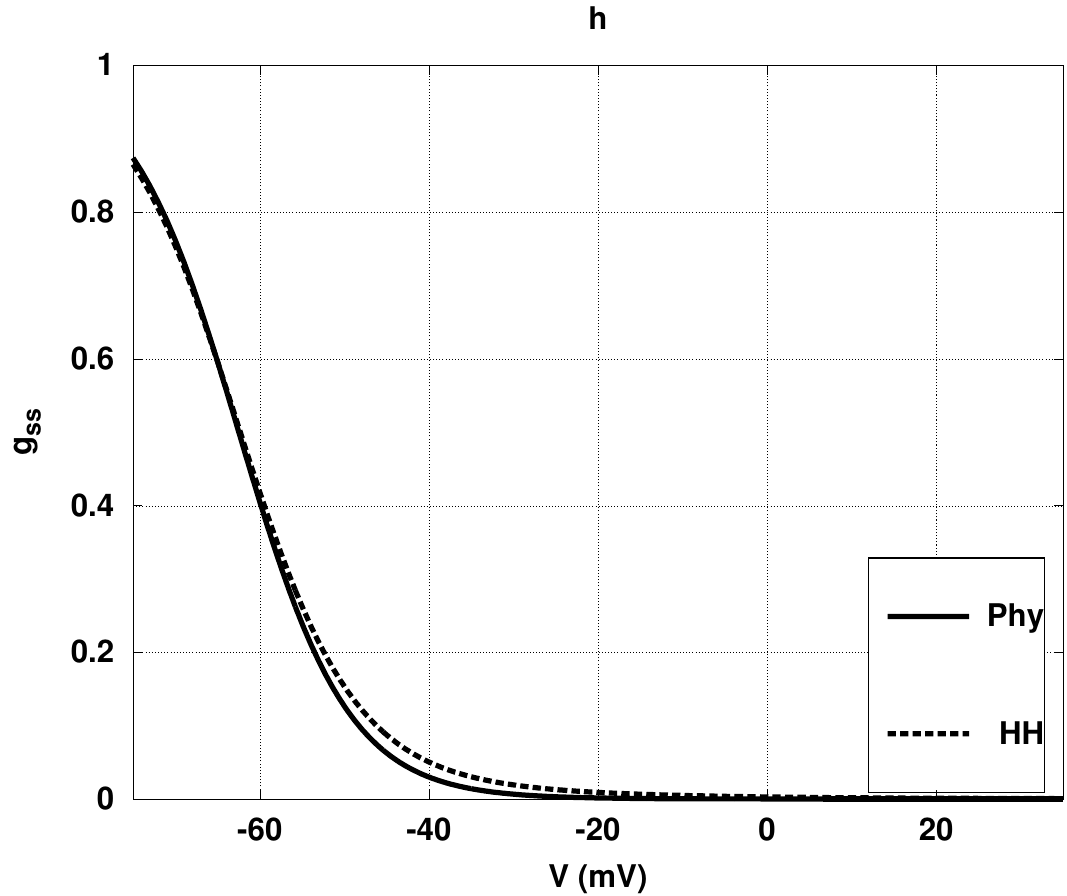}{h-gate $g_{ss}$}{0.45}
 \SubFig{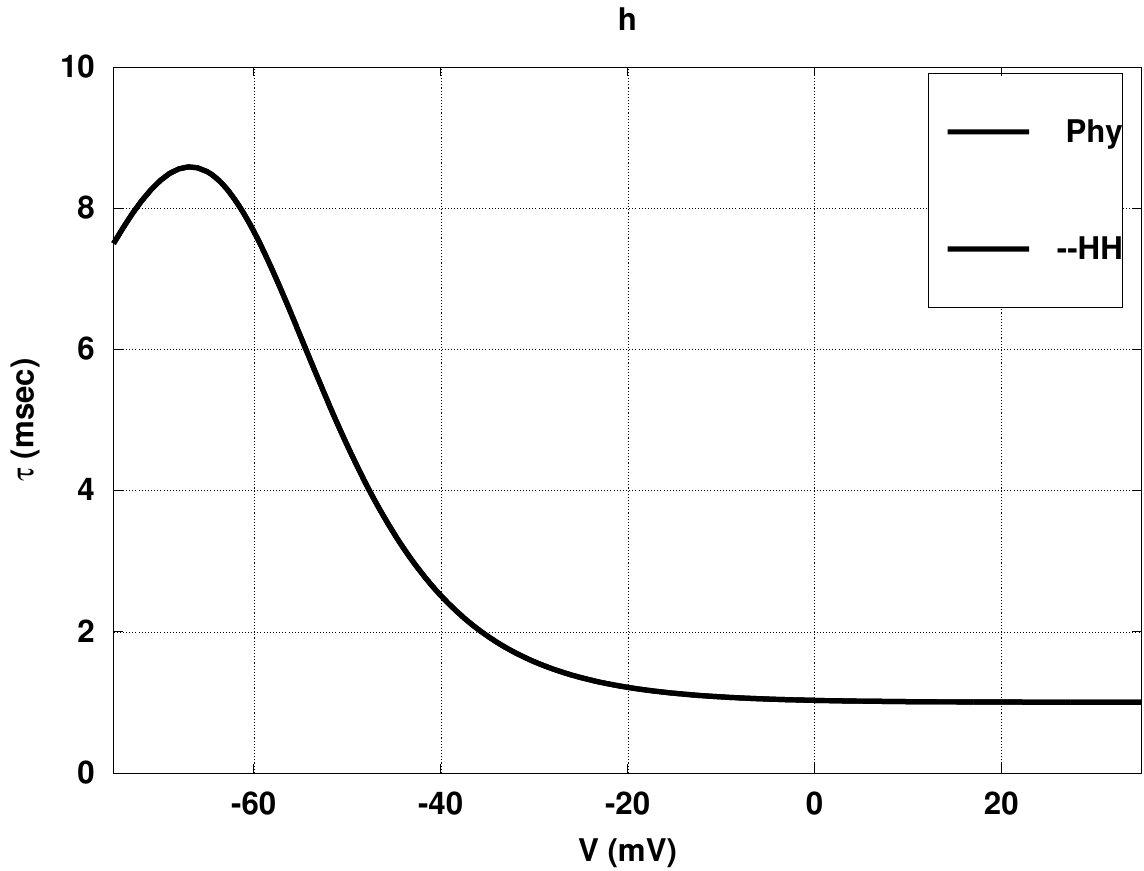}{h-gate $\tau$}{0.45}
 \caption{Physical Gate Models fitted to Hodgkin-Huxley equations.}
 \label{fig:pGate_HH}
\end{figure}
The fit is not exact, as there is no equivalence between the physical gating model and the Hodgkin Huxley empirical model. Figures \ref{subfig:Pgate_n_tau}, \ref{subfig:Pgate_m_tau} and
\ref{subfig:Pgate_h_tau} show $\tau$ for each of the three gates and
for both the physical and empirical values; $\kappa(V)$ 
has been chosen to give an exact fit by
making incorporating the empirical expressions for $\alpha$ and
$\beta$. 

%
  It is worth noting that our framework does not rely on the specific
  gating function~$G_{ion}$ chosen here for comparing an energy-based
  model approach with the HH model. 
%
Parameter values for the Bond Graph gating models are given in Table
\ref{tab:phys}. Unlike the HH model, the gates themselves draw current
from the membrane. The amount of current is partly determined by the
total gate states $x_g$. As the HH model contains no information about
gate current, the total gate states $x_g$ are chosen to give a small,
but otherwise arbitrary, value.

%
\begin{table}[htb]
 \centering
 \begin{tabular}{|l|l|l|l|}
   \hline
   Parameter & n & m & h\\
   \hline
   $z_g$ & $1$ & $3$ & $4$\\
   $k_c$ & $5.7537$ & $105.49$ & $1$\\
   $k_o$ & $1$ & $1$ & $6.3281 \times 10^{-5}$\\
   $x_g$ & $10^{-9}$ & $10^{-9}$ & $10^{-9}$\\
   \hline
 \end{tabular}
 \caption{Physical-model parameters}
 \label{tab:phys}
\end{table}

\begin{figure}[htbp]
 \centering
 \Fig{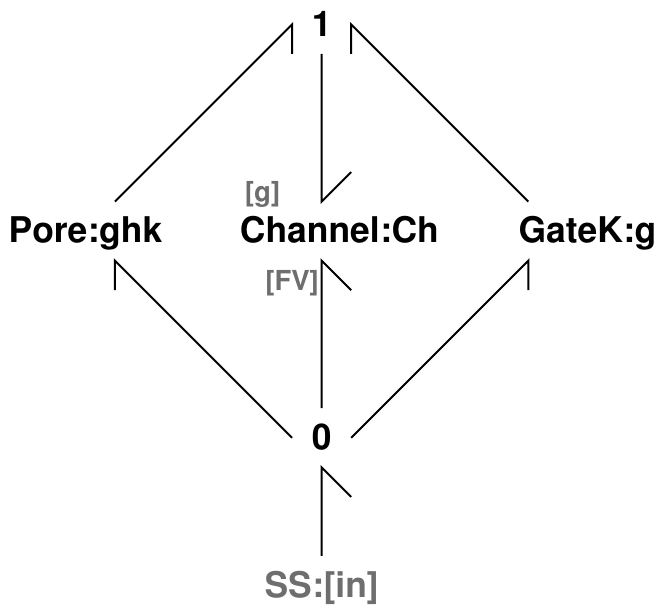}{0.3}
 \caption{Goldman-Hodgkin-Katz $\K$ Channel Model. The  $\Na$ model is
   identical except that the right-hand component is replaced by the
   appropriate gate component.  Following the Hodgkin-Huxley model, the
   leak component is not voltage gated and the right-hand component
   does not appear.
   Alternatives to the Goldman-Hodgkin-Katz model are simply obtained by
   replacing the pore component.
 }
 \label{fig:GHK_membrane}
\end{figure}

\section{Energy Flow in the Hodgkin Huxley Action Potential}
\label{sec:energy}

\begin{figure}[htbp]
 \centering
 \Fig{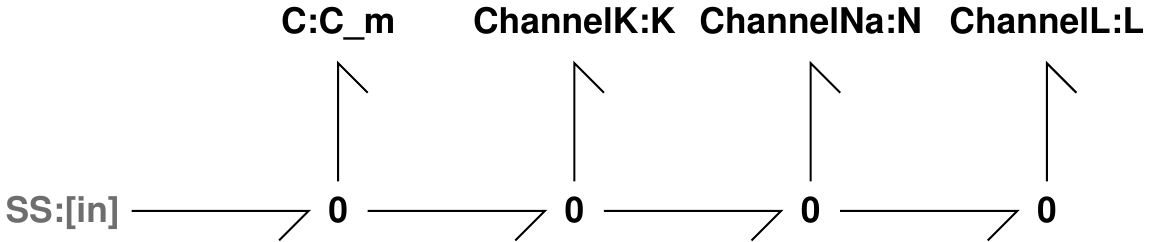}{0.6}
 \caption{Membrane model}
 \label{fig:membrane}
\end{figure}

We are now in a position to reimplement the \citet{HodHux52} model
as a physically-plausible model using the bond graph formulation. 
As mentioned in the Introduction, the Hodgkin-Huxley
model is used here as an exemplar and the general formulation of the
rest of this section is equally applicable to other more recent
models.
Following the interpretation of the model of \citet{HodHux52} given by
\citet[Chapter 5]{KeeSne09}, an area of $1\si{cm^2}$ of axon is
modelled. 
%
%
Figure \ref{fig:membrane} shows the bond graph representation of the
Hodgkin-Huxley model, consisting of three ion channels
(displayed in Figure
\ref{fig:GHK_membrane}) together with the electrogenic capacitor. These four components share the same
electrogenic potential and thus are connected to \zero junctions.
\begin{table}[htbp]
  \centering
  \begin{tabular}{|l|l|l|}
    \hline
    &  $\K$ & $\Na$ \\
    \hline
    Internal & 397 & 50\\
    External & 20 & 437\\
    \hline
  \end{tabular}
  \caption{Concentrations used in simulation (\si{mM}) \citep[Table 2.1]{KeeSne09}}
  \label{tab:conc}
\end{table}

%
The model has 13 states compared to the 4 states of the HH model. For
direct comparison with the HH model, the six states corresponding the
internal and external amounts (concentrations) of $\K$, $\Na$ and L were fixed at
constant values by applying appropriate external flows. 
Using stoichiometric analysis, the remaining conserved moieties reduce
the number of independent states to 4 -- the same as the original
Hodgkin-Huxley model. 
%
Using the pore and gate equations developed in
\S~\ref{sec:mass-action-model} and \S~\ref{sec:VoltageGating},
together with the parameters given by \citet[Chap. 5]{KeeSne09} the
reduced order system equations \citep[(3.48)]{GawCra14} were
implemented numerically.  The model flows were scaled by a factor of
$10^{-9}$ within the simulation for numerical reasons. The membrane
was initially disturbed from the resting potential by a depolarisation
of $20\si{\milli V}$.  The internal and external concentrations are
taken from \citet[Table 2.1]{KeeSne09} and are given in Table
\ref{tab:conc}. 

\begin{figure}[htbp]
 \centering
 \SubFig{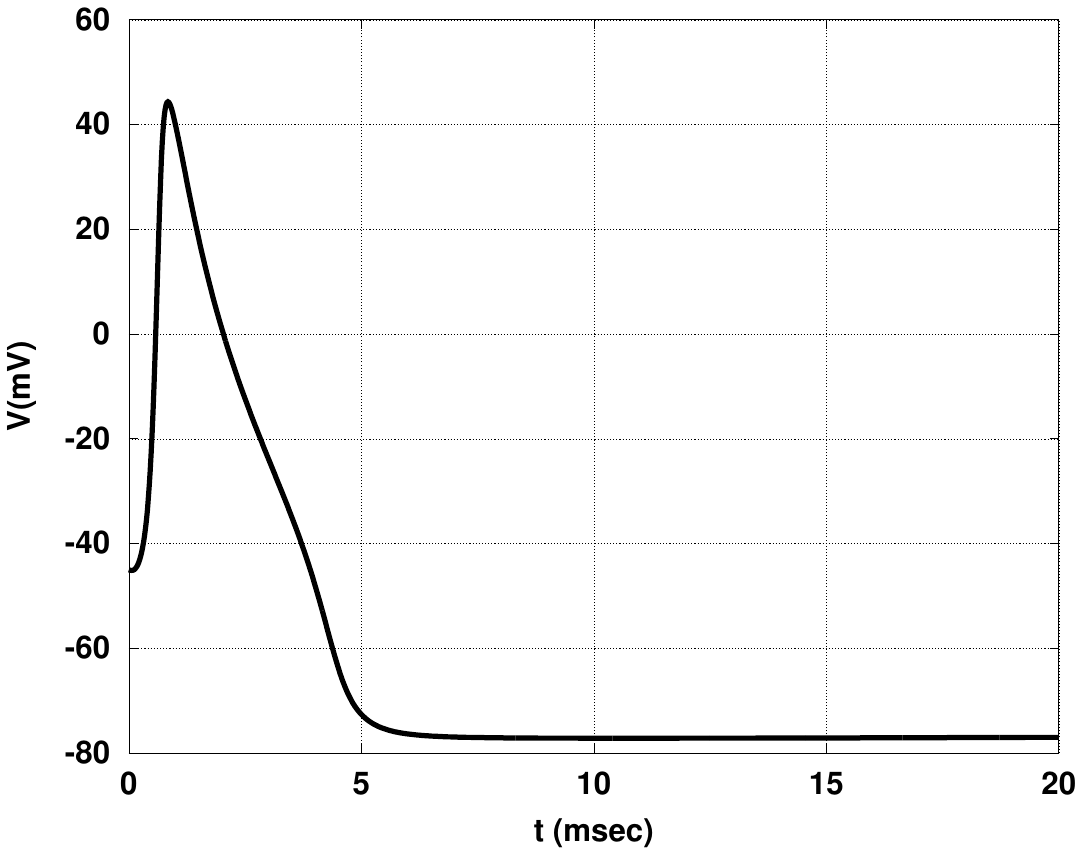}{Voltage: $V$~\si{\milli V}}{0.45}
 \SubFig{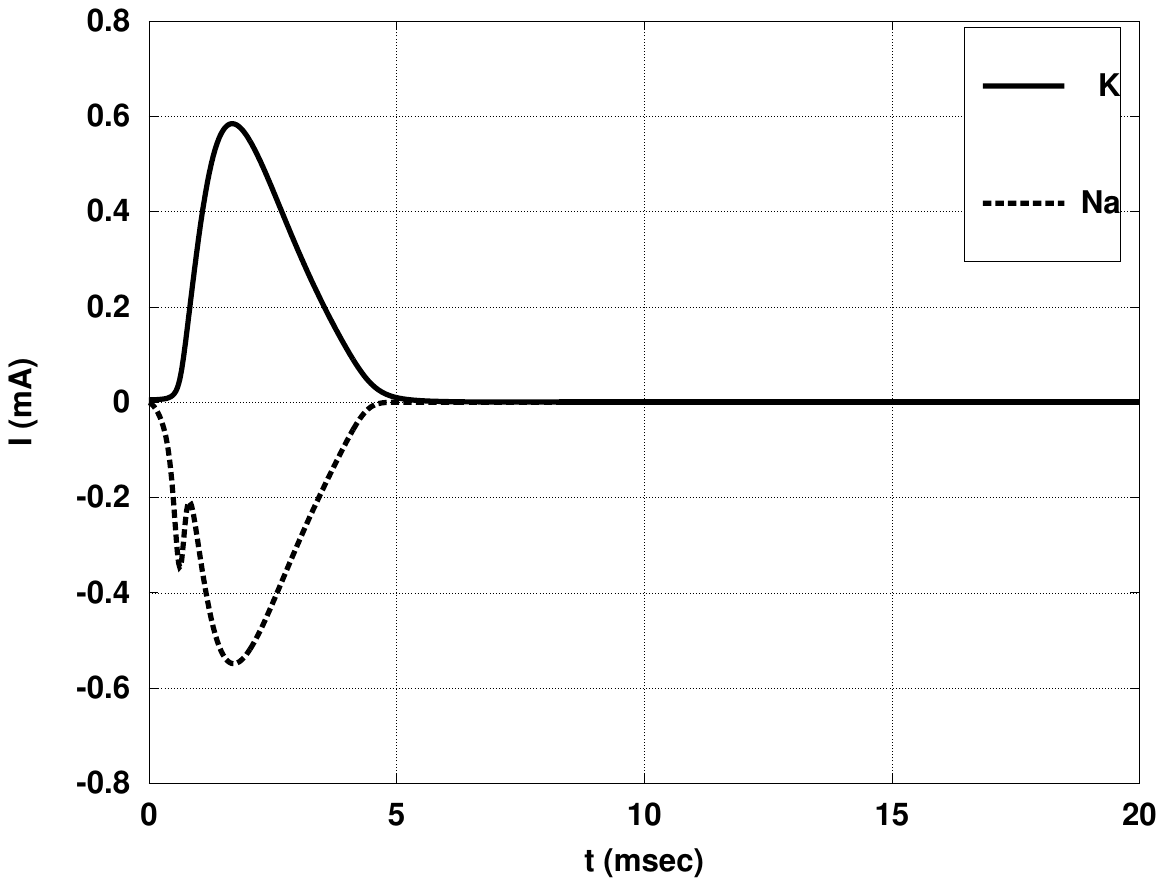}{$I_k$ \& $I_n$~\si{\micro A.cm^{-2}}}{0.45}
 \SubFig{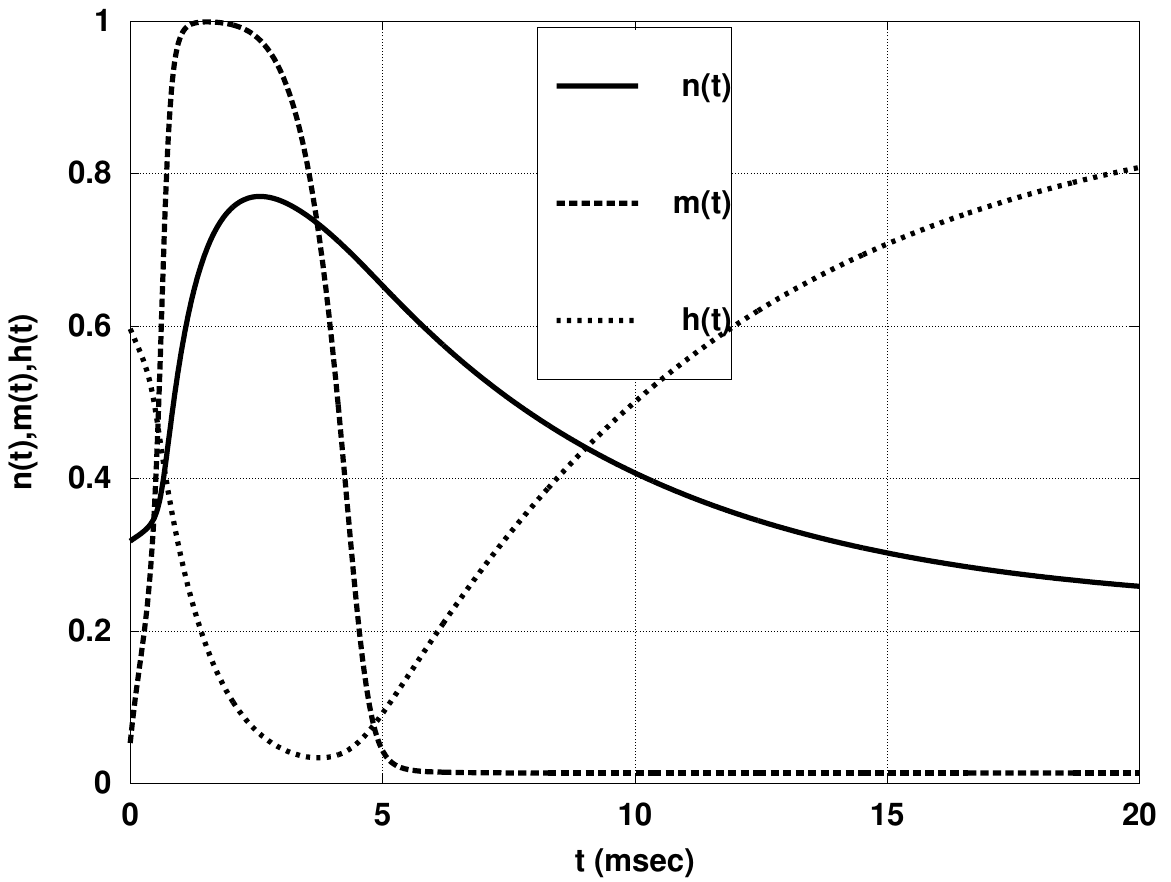}{Gating functions}{0.45}
 \SubFig{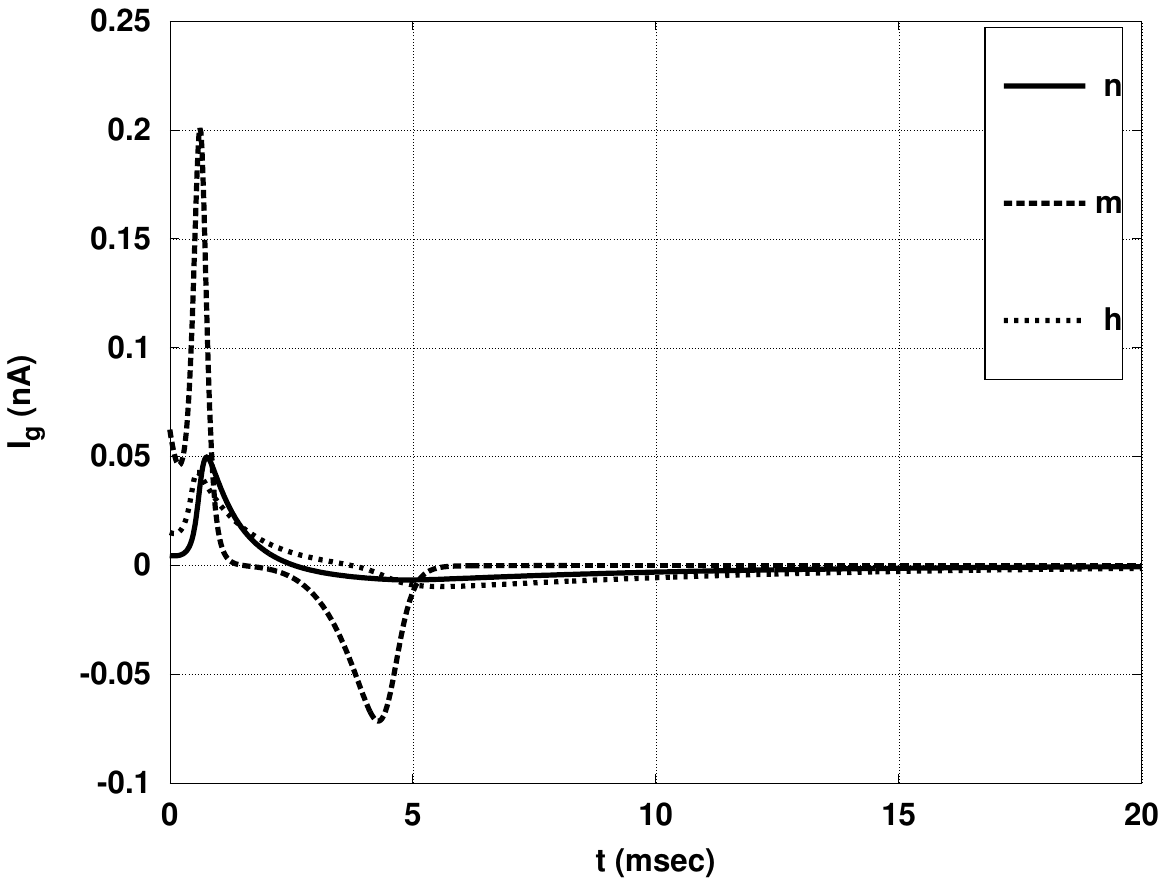}{Gate current: $I_g$~\si{\nano A.cm^{-2}}}{0.45}
   \caption{Membrane response: electrical domain. 
     (a) The membrane voltage (\si{\milli V}) is plotted against time
     (\si{msec}) for a single action potential.
     (b) The corresponding channel currents $I_k$ \& $I_n$ (\si{\micro A.cm^{-2}}).
     (c) The corresponding three gating functions $n$, $m$ and $h$.
     (d) The gate currents for the three gates (\si{\nano A.cm^{-2}}).  }
 \label{fig:membrane_resp}
\end{figure}

Figure \ref{fig:membrane_resp} shows the response in the electrical
domain. As shown by the discrepancies in Figures \ref{fig:comparison}
and \ref{fig:pGate_HH}, the physically-based models are not identical
to the original Hodgkin-Huxley model. Therefore there is a slight
discrepancy between the simulation results of Figure
\ref{fig:membrane_resp} and those shown by \citet[Chap. 5]{KeeSne09},
particularly after 5~\si{ms}.  Hence the results in the sequel
correspond to the physically-based approximation to the Hodgkin-Huxley
model rather than to the original Hodgkin-Huxley model. More
generally, the energetic consequences of different sets of parameters
for the physically-based model, or of more complex physically-based
models, could equally well be examined by the methods of this section.

%



\begin{figure}[htbp]
 \centering
 \SubFig{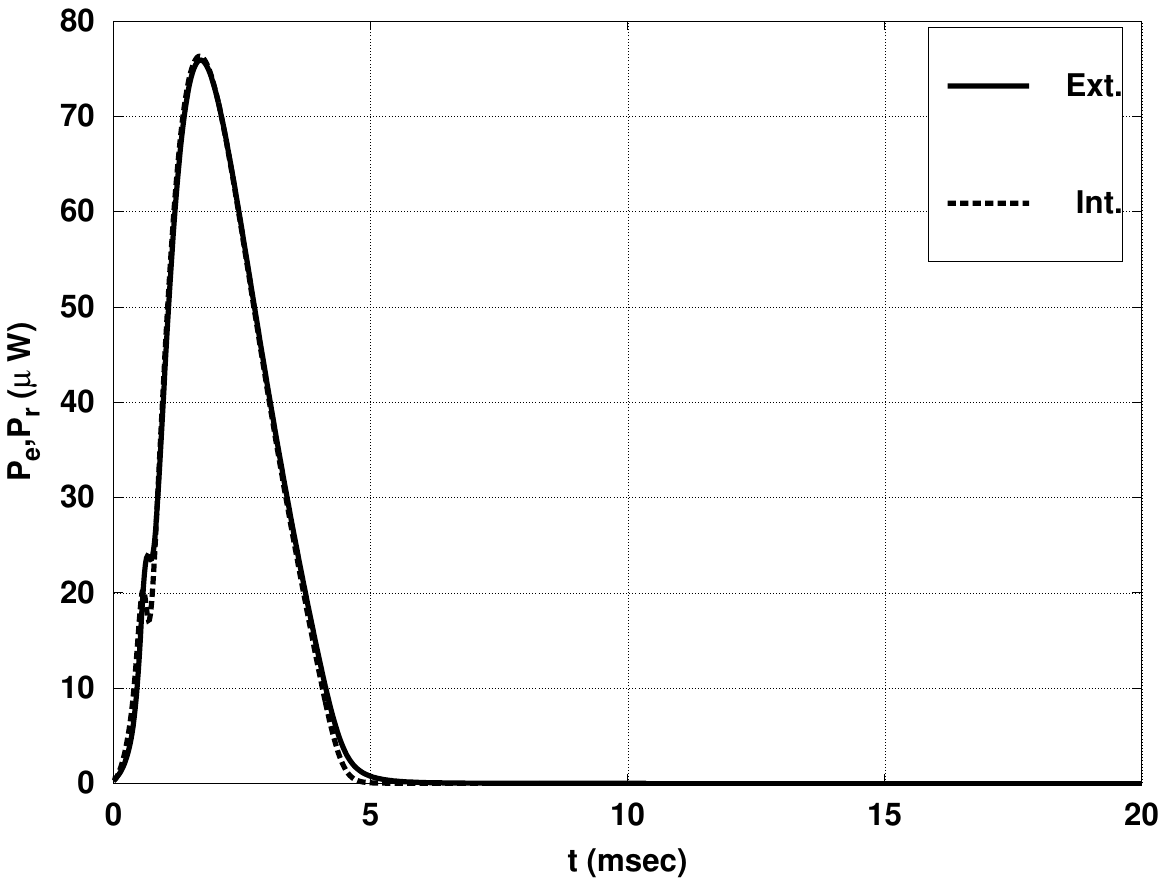}{Net power \si{\micro W.cm^{-2}}}{0.45}
 \SubFig{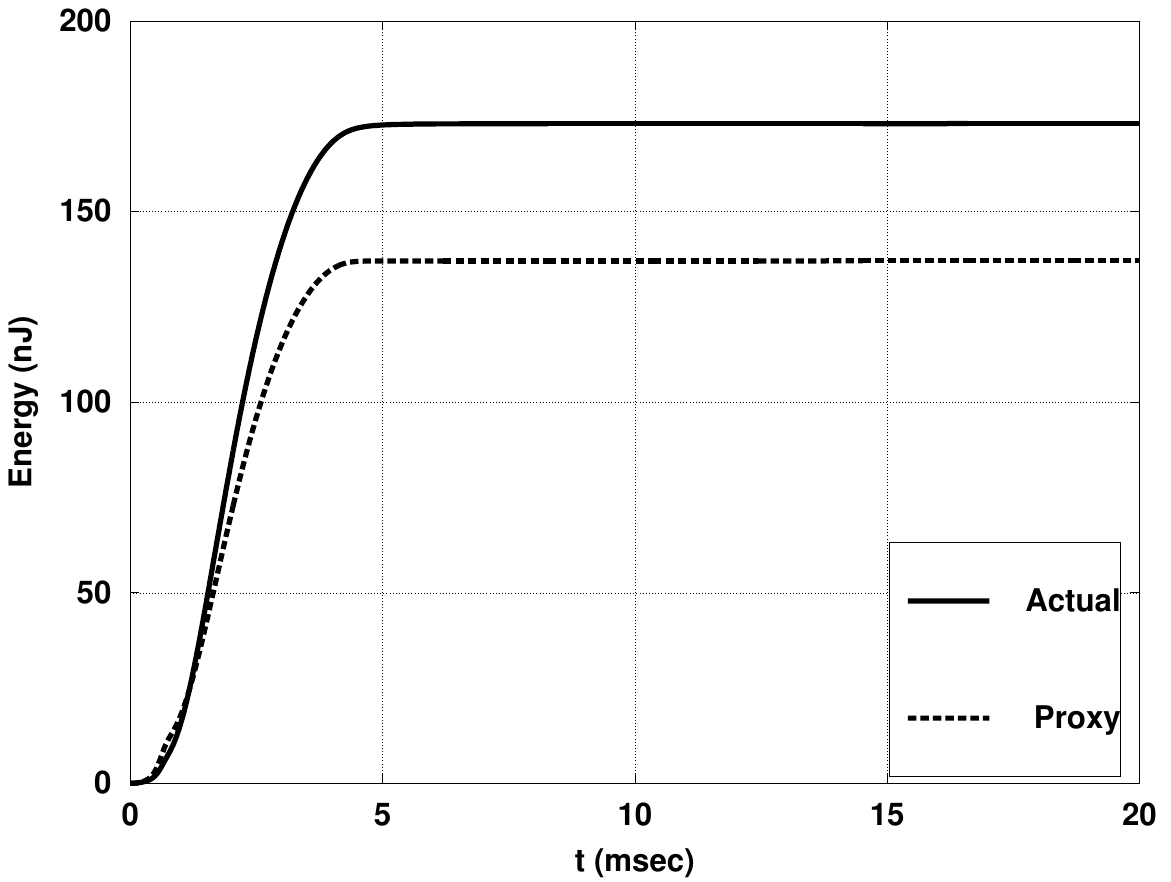}{Energy \si{nJ.cm^{-2}}}{0.45}
   \caption{ Power \& Energy plotted against time (\si{msec}).
     (a)~The net external power and dissipated power (\si{\micro
       W.cm^{-2}}).
     %
     %
     (b)~The external energy (time integral of
     external power) and the ATP-proxy energy (\si{nJ.cm^{-2}}). 
%
}
\label{fig:membrane_resp_energy}
\end{figure}
%
%

In order to calculate the energy flows associated with these ionic
movements, \citet{GawCurCra15} give formulae for the energy flows in
the bond graph of biochemical networks
. As noted above, the rate of energy transfer, or power flow, of a
bond can be calculated as the product of effort and flow. Thus, the
power flow associated with substance $A$ is $p_A = \mu_A v_A \si{~W}$
where $\mu_A$ is the chemical potential defined in Equation
\eqref{eq:mu_A_0} and $v_A$ is the molar flow rate. In the particular
case relevant here that substance $A$ occurs on each side of a
membrane and is replaced on one side, and removed at the other, at a
variable rate $v_A$ so that it remains at a fixed concentration on
each side, the net external power associated with substance $A$ is:
\begin{equation}\label{eq:P_ea}
  p_{eA} = \lb \mu_i - \mu_r \rb v_A
\end{equation}
where $\mu_i$ and $\mu_r$  are the internal and external chemical
potentials respectively. Using Equation
\eqref{eq:mu_A_0}
\begin{align}
  \mu_i - \mu_r &= RT \lb \ln \mf_i - \ln \mf_r \rb 
  = RT  \ln \frac{\mf_i}{\mf_r} \notag\\
 &=  RT  \ln \frac{c_i}{c_r} = G_A \label{eq:G_A}
\end{align}
where $c_i$ and $c_r$ are the internal and external concentrations of
substance $A$; $G_A$ can be interpreted as the Gibbs free energy change of
moving $A$ from inside to outside in this particular case.

In the case of the $i$th chemical reaction component, the energy inflow  is
the product of the reaction flow $v_i$ and the forward affinity
$A^f_i$ of Equation \eqref{eq:A^f} and the energy flow out is
the product of the reaction flow $v_i$ and the reverse affinity
$A^r_i$ of Equation \eqref{eq:A^r}. Thus the power dissipated in the $i$th chemical reaction is:
\begin{equation}\label{eq:p_i}
  p_i = A^f_i v_i - A^r_i v_i = A_i v_i
\end{equation}
where the reaction affinity $A_i$ is given by $A_i = A^f_i-A^r_i$.

Using the same simulation data as Figure \ref{fig:membrane_resp},
Figure \ref{subfig:Memb_P_er} shows the external power
($P_e~\si{\micro W.cm^{-2}}$) due to the flows of \si{K^+}and \si{Na^+}
required to keep the concentrations constant and the net rate of
energy dissipation ($P_r~\si{\micro W.cm^{-2}}$). These two powers are
approximately the same; the difference is due to transient energy
storage in the electrogenic capacitor.
%
%
The corresponding energy $E_e(t)$ of an action potential is computed
by integrating the external power $P_e$ with respect to time
\begin{equation}
  E_e(t) = \int_0^{t} P_e(\tau) d\tau
\end{equation}
$E_e(t)$ is plotted in Figure \ref{subfig:Memb_ATP} with the legend
``Actual''; the total energy required for this particular model is
about $173\si{~nJ.cm^{-2}}$.
\footnote{
    The energy corresponding to the initial depolarisation of
    $V_0=20\si{\milli V}$ is $\frac{1}{2}C_mV_0^2 = 0.2\si{nJ.cm^{-2}}$; this can be
    neglected in the overall energy balance.
}

It is interesting to compare this precise method of computing
energy dissipation with the ATP proxy approach. The influx of $\Na$
\citep{CarBea09,HasOttCal10,SenSteLau10,SenSte14} is often taken as a
proxy for energy consumption. For example, as discussed by
\citet{SmiCra04} and \citet{HasOttCal10}, $3\Na$ ions are moved back
across the membrane using 1 ATP molecule by the sodium-potassium pump
($\Na$, $\K$, ATPase). 
  Thus the total $\Na$ passing though the membrane during an action
  potential $x_n$ can be taken as a proxy for energy consumption.  For
  example, \citet{HasOttCal10} use the \emph{ATP-proxy} $x_a$ via the
  formula $x_a = \frac{x_n}{3}$.
  The energy corresponding to the ATP-proxy $x_a$ can be computed from
  \begin{equation}
    E_a = G_{ATP} x_a
  \end{equation}
  where $G_{ATP}\approx 31 \si{~kJ.mol^{-1}}$ is the Gibbs free energy
  associated with the reaction $ATP \reac ADP + P_i$.
  $E_a$ is plotted in Figure \ref{subfig:Memb_ATP} together with the
  actual energy $E_e$; the ATP-proxy energy required for this
  particular model is about $137\si{~nJ.cm^{-2}}$ -- a discrepancy of
  about $20\%$ which is discussed in the sequel.
  Alternatively, this energy requirement could be reexpressed in units
  of moles of $ATP$ by dividing $E_e$ and $E_a$ by $G_{ATP}$. 

  Different physically-based action potential models will have
  different discrepancies between the ATP proxy approach and the
  actual energy. Thus comparing action potential energy requirements
  of different models (whether physically-based or
  non-physically-based) using the ATP proxy approach may be
  misleading.



\section{Energy consumption in healthy and degenerative retinal ganglion cells}
\begin{figure}[htbp]
  \centering
    \SubFig{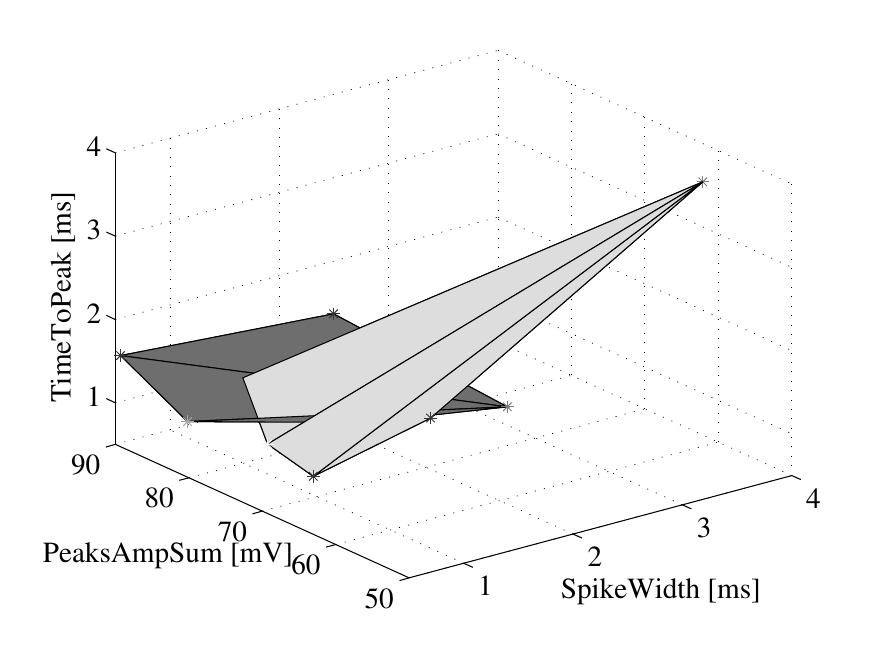}{Experimental data}{0.45}
    \SubFig{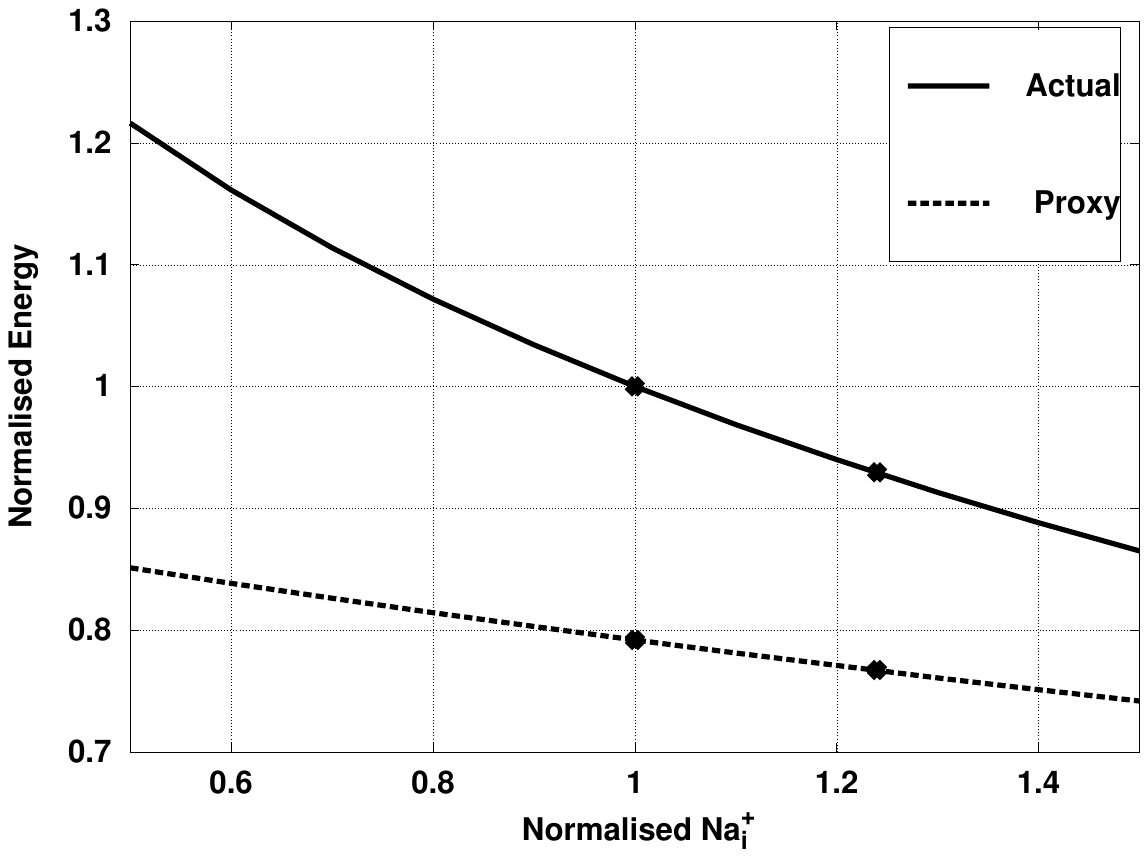}{Simulated energy consumption}{0.45}
    \caption{Retinal Ganglion Cells.
(a) Topological manifolds for the parameters of action potentials in
WT and RD1 mice. Pale: WT mice; dark: RD1 mice. TimeToPeak is
calculated as a difference between the time of the maximum amplitude
of an action potential and the time taken for the membrane potential
to reach $dV/dt>10$ mV/ms threshold. PeaksAmpSum is calculated as a sum of the maximum amplitude of an action potential and absolute hyperpolarisation level.
(b) Simulated Energy consumption as internal $\Na$ concentration
varies: the ratio of the energy consumption to the actual energy
consumption at the nominal internal $\Na_i$ concentration $\Na_0$ is
plotted against $\Na_i/\Na_0$ for both actual and proxy energy
consumption.}
  \label{fig:vary}
\end{figure}
\begin{table}[htbp]
  \centering
  \begin{tabular}{|l| c c | c |}
    \hline
    & Experiments & &Simulations\\
    \hline
    \textbf{WT mice} & mean & std &\\
    Vmax [mV] & 23 &6 & 23 \\
    Spike Width [ms] &1.9 &0.8 & 2 \\
    \hline
    \textbf{RD1 mice} & mean & std &\\
    Vmax [mV] & 29 & 9 & 28 \\
    Spike Width [ms] &1.5 &0.5 & 1.6 \\
    \hline
  \end{tabular}
  \caption{Comparison of the experimental data (n=8 for WT and n=6 for
    RD cells) and simulation results. Mean values and standard errors are
    given for the experiments. Results are shown using standard ionic model formulation. 
  }
  \label{tab:comp}
\end{table}
As discussed in Appendix \ref{sec:appl-prop-meth}, \emph{in vitro}
data was collected and analysed from retinal ganglion cells (RGCs) of
wild-type (WT) and degenerative (RD1) mice.  Figure
\ref{subfig:ExperData} illustrates that the experimentally recorded
action potentials belong to separate topological manifolds for WT and
RD1 mice. This suggests that energy consumption for single action
potentials is different for WT and RD1 mice.

The methodology employed here to account for differences in the shape
of action potential in WT and RD1 cells assumes that sodium and
potassium conductances contribute the most to the spike width and
spike maximum amplitude.  Other conductances may also contribute to
the shape of the action potential; however, their effect was not
investigated here.  Due to sodium conductance contributing maximally
to the amplitude of action potentials, this conductance may be
overestimated for RD1 cells that have larger mean spike height. This
does not affect the qualitative results presented here, but is a
possible limitation for the interpretation of quantitative results.

  Using experimentally fitted parameters as described in the
  Supporting Material, simulations of both cell types were conducted
  and a comparison of the experimental data and simulation results is
  given in Table \ref{tab:comp}.

  The simulation results indicated that $\bar{g}_{\mathrm{Na}}$ is
  increased by at least 24$\%$ in degenerative retina.  Figure
  \ref{subfig:Memb_vary} shows how actual and ATP-proxy energy varies
  with internal sodium concentration. The solid line illustrates
  actual energy consumption, the dashed line illustrates the ATP-proxy
  calculation, and the crosses indicate data points calculated for RD1
  and WT mice energy consumption. Figure \ref{subfig:Memb_vary}
  illustrates that the difference in the normalised energy consumption
  for one action potential between WT and RD1 RGCs is 0.03 using
  ATP-proxy methodology, while the difference is 0.08 using
  methodology proposed in this paper. The figure shows that the
  ATP-proxy methodology underestimates the energy consumption for both
  mouse types. In addition, the ATP-proxy methodology underestimates
  the difference between the two types of mice, while the proposed
  methodology provides accurate comparison of energy consumption
  between two similar-shaped action potentials (note the relatively
  small differences between WT and RD1 cells in Table \ref{tab:comp}).

\section{DISCUSSION}
\label{sec:conclusion}
\begin{table}[htbp]
  \centering
  \begin{tabular}{|l|l|l|l|l|}
    \hline
    & $\K$ & $\Na$ & $\K+\Na$ &ATP\\
    \hline
    x (\si{pmol}) & 13.60 & 13.27  &--& 4.42\\
    G (\si{kJ.mol^{-1}}) & 7.54 & 3.68 &--& 31\\
    E (\si{nJ}) & 101.35 & 71.73 &173.08 & 137.08\\
    \hline
  \end{tabular}
  \caption{Simulation results (per \si{cm^{2}}). $x$ is the quantity,
    $G$ the molar free energy and $E$ the energy consumed by the
    action potential. $x_{a}$ is computed from the formula $x_a = \frac{x_n}{3}$}
  \label{tab:res}
\end{table}
We have extended our earlier work on energy-based bond graph modelling
of biochemical systems to encompass chemoelectrical systems. In
particular, we have introduced the electrogenic capacitor and the
electrostoichiometric transformer to bridge the chemical and
electrical domains.

As a particular example illustrating the general approach, we have
constructed a bond graph model of the \citet{HodHux52} model of the
axon; and we have used this model to show that calculation of energy
consumption during generation of the action potential by counting
$\Na$ ions crossing the membrane underestimates true energy
consumption by around 20\%.

  In this particular situation, the concentrations are constant and
  thus Equation \eqref{eq:G_A} is appropriate. Moreover, the
  contribution of the leakage and the gating currents are small and
  can be neglected. The values of Table \ref{tab:conc} and Equation
  \eqref{eq:G_A} gives the molar free energy values of Table
  \ref{tab:res} for $\K$, $\Na$; that for ATP is taken from
  \citet[(1.23)]{KeeSne09}. Because the actual energy consumption
  depends on both the amounts of $\K$, $\Na$ as well as on the
  internal and external concentrations, there is no way that the
  ATP-proxy formula (based on only the amount of $\Na$) can give the
  correct value under all circumstances. To illustrate this, Figure
  \ref{fig:vary} shows how actual and ATP-proxy energy varies with
  internal $\Na$ concentration expressed as a ratio $\rho$ to the
  nominal concentration of Table \ref{tab:conc}. The discrepancy
  between actual and ATP-proxy energy varies with $\rho$.
  Moreover, the method of this paper does \emph{not} require the
  concentrations to be constant during an action potential and is thus
  applicable to more general situations.



The wider significance of our approach is that it provides a framework
within which biophysically based models are robustly thermodynamically
compliant \citep{GawCurCra15}, as required for example when
considering the energetic costs and consequences of cellular
biological processes. Furthermore, the Bond Graph approach provides a
basis for modular modelling of large, multi-domain electro-chemical
biological systems, such as is now commonplace in systems biology
models of excitable membranes in the neuronal and cardiac
contexts. Components and modules which are represented as Bond Graphs
are physically plausible models which obey the basic principles of
thermodynamics, and therefore larger models constructed from such
modules will also consequentially be physically plausible
models. Future work will further develop these concepts in order to
represent ligand-gated ion channels, ion pumps (such as the $\Ca$ pump
SERCA \citep{TraSmiLoiCra09,GawCra14} and the $\Na$ pump
\citep{SmiCra04,TerNieCra08}, as are required for current generation
neuronal and cardiac cell models.
  This modular approach allows simpler modules to be replaced by more
  complex modules, or empirical modules to be replaced by
  physically-based modules, as the underlying science advances.

Furthermore, the multi-domain nature of Bond Graphs makes possible
extension of the approach to mechano-chemical transduction. In
actively contracting cardiac muscle, for example, energetic
considerations are dominated by force production, where approximately
75--80\% of ATP consumption in cardiomyocytes over a cardiac cycle is
due to formation of contractile cross-bridges, 5--10\% due to the
$\Na$ pump, and $\Ca$ extrusion and uptake into $\Ca$ stores
accounting for the remainder \citep{TraloiCra15}. In cardiac muscle,
energetics is known to play a critical role in the health of cardiac
muscle, with many studies implicating energetic imbalance or
inadequacy of energy production in heart disease \citep{Neu07}. Models
which provide a mechanism with which to assess the energetic aspects
of cell function are therefore much needed. Combining metabolism,
electro-chemical and chemo-mechanical energy transduction to examine
energy flows within the heart \citep{Neu07,Kat11} is therefore a major
goal of our work.
Conservation of mass and energy have been used by \citet{WeiUllSch14}
and \citet{UllWeiDah15} to examine the dynamics of seizure and
spreading depression. It would be interesting to reexamine this work
in the more general context of bond graph modelling.

The ATP-proxy approach is based on assuming that the biological entity
is operating in a normal state and therefore could lead to misleading
conclusions in a pathophysiological state. In contrast, our approach
makes no assumption of normality and may be expected to be of use in
pathophysiological states in general and, in particular, the retinal
example discussed in this paper.
Our use of the RD1 degenerate retina mouse model ensures that the outcomes
of this project are directly relevant to human patients since RD1 mice
have a degenerate retina that has distinct similarities to that
observed in human patients with \emph{retinitis pigmentosa} -- a set
of hereditary retinal diseases that results from the degenerative loss
of the photoreceptors in the retina. It has been proposed that the
death of rod photoreceptors results in decreased oxygen consumption
\cite{SheYanDon05}. In addition, it has been shown that potassium
channel-opening agents directly affect mitochondria
\cite{KulKudSze08}.  Therefore, it is important to understand how
energy consumption in degenerative retina is altered.  The proposed
methodology allows a comparison between the energy consumption in
healthy and degenerate mice, even when the differences in action
potentials between the two types are small.

  The modularity of the bond graph approach allows the action
  potential models of this paper to be combined with models of the
  various trans-membrane pumps and transporters to give a
  thermodynamically correct model of ATP consumption. It would
  therefore be interesting to reexamine the optimality arguments of
  \citet{HasOttCal10} and \citet{SenSte14} using this approach. 

  As shown by \citet{Mit76,Mit11}, the key feature of mitochondria is
  \emph{chemiosmotic} energy transduction whereby a chain of redox
  reactions pumps protons across the mitochondrial inner membrane to
  generate the \emph{proton-motive force} (PMF). This PMF is then used
  to power the production of ATP -- the universal fuel of living
  systems. The ideas expressed in this paper have been recently
  extended to model chemiosmotic energy
  transduction~\citep{Gaw17a}. 
  Future work will combine the chemoelectrical bond graph models of
  this paper with bond graph models of anaerobic metabolism
  \citep{GawCurCra15}, and bond graph models of mitochondrial chemiosmotic energy
  transduction~\citep{Gaw17a}, to give integrated models of
  neuronal energy transduction suitable for investigating neuronal
  dysfunctions such as Parkinson's disease
  \citep{WelClo10,Wel12,WelClo12}.

\section{ACKNOWLEDGEMENTS}
We would like to thank the reviewers for their suggestions for
improving the paper.
P.G. would like to thank the Melbourne School of Engineering
for its support via a Professorial Fellowship. 
I.S. gratefully acknowledges funding from Sonderforschungsbereich
(SFB) 803, project Z02 (Deutsche Forschungsgemeinschaft, Germany).
This research was in part conducted and funded by the Australian
Research Council Centre of Excellence in Convergent Bio-Nano Science
and Technology (project number CE140100036).

\section{AUTHOR CONTRIBUTIONS}
The theory was developed by PJG, IS and EC. The theoretical part of
the paper was written by PJG, IS and EC and the experimental part by
TK. Experiments were conducted by SS and MI.

\section{DATA ACCESSIBILITY}
A virtual reference environment \citep{HurBudCra14} is available for
this paper at \url{https://github.com/uomsystemsbiology/energetic_cost_reference_environment}.

\newpage
\appendix
\section{Application of the proposed methodology to calculate the
  energy consumption in healthy and degenerative retinal ganglion
  cells}
\label{sec:appl-prop-meth}
\emph{In vitro} data was collected and analysed from retinal ganglion
cells (RGCs) of wild-type (WT) (n=8) and degenerative RD1 (n=6) mice 4
- 4.5 month old. Experimental procedures were approved by the animal
welfare committee at the University of Melbourne and are in accordance
with local and national guidelines for animal care. Animals were
housed in temperature-regulated facilities on a 12h light/dark cycle
in the animal house and had plentiful access to food and
water. Neither WT nor RD1 mice were dark adapted for these
experiments.

Retinae from WT and RD1 mice were treated identically. Mice were
anaesthetised with simultaneous ketamine (67 mg/kg) and xylazine (13
mg/kg) injections, the eyes were enucleated and then the mice were
killed by cervical dislocation. Their eyes were bathed in carbogenated
(95$\%$ O2 and 5$\%$ CO2) Ames' medium (Sigma-Aldrich, St. Louis, MO),
hemisected at the ora serata, and the cornea and lens were
removed. The retina was continuously superfused with carbogenated Ames
medium at a rate of 4-8 ml/min. All of the procedures were performed
at room temperature and in normal room light.

The flat-mount retina was viewed through the microscope with the use
of Nomarski DIC optics and also on a video monitor with additional 4x
magnification using a CCD camera (Ikegami, ICD-48E). For whole cell
recording, a small opening was first made with a sharp tip pipette
(resistance above 14 M$\Omega$) through the inner limiting membrane
and optic fibre layer that covered a selected retinal ganglion
cell. Prior to recording, the pipette voltage in the bath was
nullified. The pipette series resistance was measured and compensated
for using standard amplifier circuitry (SEC-05x; NPI Electronic
Instruments). Pipette resistance was in the range of 7-14 M$\Omega$
for all experiments.  Membrane potentials were amplified (as above
with SEC-05x, npi) and digitised at 50 kHz (USB-6221, National
Instruments), acquired and stored in digital form by custom software
developed in Matlab (Mathworks).





In this study, we sought to account for the differences in the energy
consumption between WT and RD1 RGCs on the basis of differences in the
magnitudes of the maximal conductance of sodium and potassium
currents, $\bar{g}_{\mathrm{Na}}$ and $\bar{g}_{\mathrm{K}}$
respectively.  While all other parameters were kept fixed,
$\bar{g}_{\mathrm{Na}}$ and $\bar{g}_{\mathrm{K}}$ were systematically
varied in the range [$10^{-15},0.1$] in variable steps (higher
resolution for smaller values).  For conservative calculation of the
difference in the energy consumption, the smallest difference between
$\bar{g}_{\mathrm{Na}}$ and $\bar{g}_{\mathrm{K}}$ in WT and RD1 types
that replicated experimental data is reported here. Model parameters
used in simulations are given in Table 5; $\bar{g}_{\mathrm{i}}$,
$V_{\mathrm{i}}$ are maximum conductance and reversal potential of the
current ``i''. $V_{\mathrm{Ca}}$ and $\bar{g}_{\mathrm{K(Ca)}}$ depend
on the intracellular Ca$^{2+}$ concentration. 
  At the moment, no experimental data is available on the properties of
  gating parameters for RGS in RD1 mice.  Due to this, the same gating
  parameters were used to model action potential in healthy and
  degenerative tissue. 
A single-compartment Hodgkin-Huxley type neurons was simulated in
NEURON. The standard Euler integration method was used in simulations.
Data was analysed in Matlab (Mathworks).

\begin{tabular}{l l}
\multicolumn{2}{l}{Table 5. Simulation parameters. }\\
\hline
$T=22^0 \;$ C & $C_m=1 \; \mu$ F/cm$^2$ \\
$V_{\mathrm{Na}}=35 \;$ mV & $\bar{g}_{\mathrm{Na}}$  is varied in simulations\\
$V_{\mathrm{Ca}} \;$ is variable, refer to \cite{KamMefBur12} & $\bar{g}_{\mathrm{Ca}}=2.2 \cdot 10^{-3}\;$ S/cm$^2$\\
$V_{\mathrm{K}}=-70 \;$ mV & $\bar{g}_{\mathrm{K}}$ is varied in simulations\\
    &               $\bar{g}_{\mathrm{K,A}}=3.6 \cdot 10^{-2} \;$ S/cm$^2$\\
      &              $\bar{g}_{\mathrm{K(Ca)}}$ is variable, refer to \cite{KamMefBur12}\\
$V_{\mathrm{L}}=-60 \;$ mV &  $ \bar{g}_{\mathrm{L}}= 10^{-6} \;$ S/cm$^2$\\
$V_{\mathrm{h}}=0 \;$ mV &  $\bar{g}_{\mathrm{h}}=10^{-7}$ S/cm$^2$ \\
$V_{\mathrm{T}}=120 \;$ mV & $\bar{g}_{\mathrm{T}}=10^{-3}$ S/cm$^2$\\
\hline
\end{tabular}
 The following simulation parameters reproduced the maximum amplitude
  and width of the experimentally recorded action potentials in WT and
  RD1 mice and gave the smallest difference in values between the two
  retina types.  Parameters for WT: $\bar{g}_{\mathrm{Na}}=0.0342$
  S/cm$^2$, $\bar{g}_{\mathrm{K}}=0.0102$ S/cm$^2$. Parameters for
  RD1: $\bar{g}_{\mathrm{Na}}=0.0422$ S/cm$^2$,
  $\bar{g}_{\mathrm{K}}=0.0102$ S/cm$^2$.

\end{document}